\documentclass[%
 reprint,
prb,
floatfix,
]{revtex4-2}

\usepackage{blindtext}
\usepackage{graphicx}
\usepackage{dcolumn}
\usepackage{bm}
\usepackage{amsmath}
\usepackage{booktabs}
\usepackage{tabularx}
\usepackage{appendix}
\usepackage{footmisc}
\begin{document}

\preprint{APS/123-QED}

\title{Evidence of Ordering in Cu-Ni Alloys from Experimental Electronic Entropy Measurements }

\author{\textit{Jonathan Paras, Antoine Allanore}}
\affiliation{Department of Materials Science and Engineering,\\Massachusetts Institute of Technology, Cambridge, MA, USA }

\date{\today}

\begin{abstract}
Phase diagrams exhibiting extended solid-solution and lens-like melting are often reproduced using ideal solutions, where ideal mixing considers a fully random configurational entropy of mixing. In the field of irreversible thermodynamics, experimental measurements of the composition variation of high-temperature electronic transport and molten-state properties suggest however a strong role for short-range atomic ordering in these systems. Herein, measurements of the thermopower and resistivity are reported for Cu-Ni solid-solutions as a function of temperature and composition. The electronic transport properties were interpreted with an irreversible thermodynamic framework, revealing a large electronic contribution to the entropy of mixing. Through appeal to a cluster model for the configurational entropy that uses the electronic contribution to inform the existence of ordered associates, we rationalize such contribution of the electronic entropy with the ideal entropy of mixing commonly used to model such systems. These results suggest that the short range order (S.R.O.) of the atoms plays a significant role in both the solid and molten states, even when there are no dominant intermetallic compounds in these alloys. 
\end{abstract}

\maketitle

\section{Introduction}

\begin{figure}[h]
	\includegraphics[width=\linewidth]{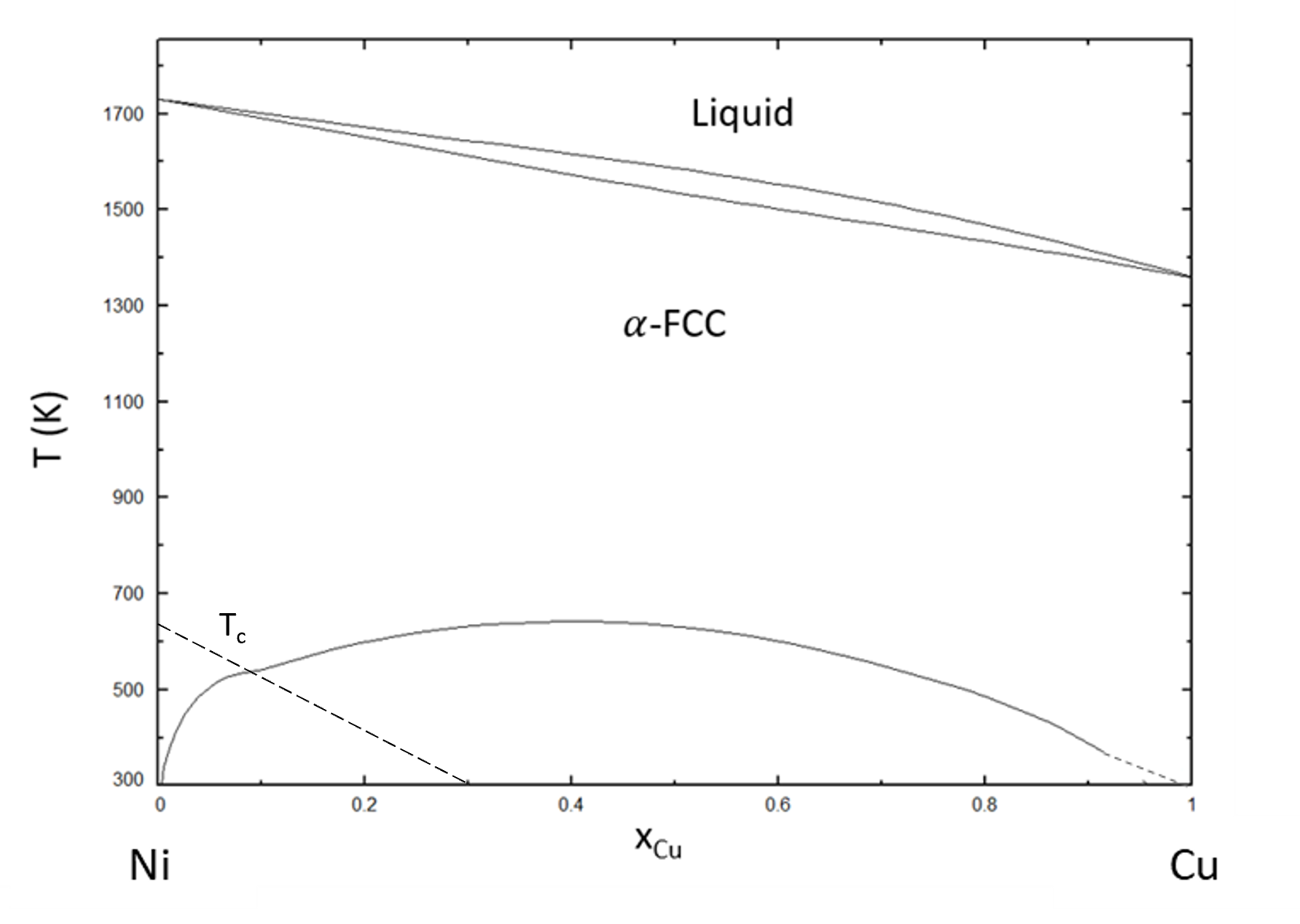}
				\caption{Cu-Ni phase diagram computed with FactSage 8.0 using the FS Steel Database \cite{sundman1985thermo}, suggesting full miscibility in the solid state (phase $\alpha$ FCC) at high temperature and the liquid state, with a possible low temperature miscibility gap. Curie data from Turchanin et al
    \cite{Turchanin2007}}
    \label{CuNiPhaseDiagram}
				\end{figure}

Cu-Ni, which phase diagram is illustrated in Figure (\ref{CuNiPhaseDiagram}), forms a continuous solid solution alloy across the entire binary composition range. The liquidus and solidus adopt the shape of a lens, and by inspection, exhibit no remarkable features otherwise \cite{hultgren1973selected,hansen1958constitution}. Whereas the solid-solution has been treated as if it were thermodynamically ideal, the liquid state properties of Cu-Ni alloys, including the surface tension, viscosity, and density, have been demonstrated to be far from ideal, in some cases exhibiting multiple changes in concavity as the composition changes across the binary, including as the liquid is superheated \cite{prasad1991surface,chikova2019viscosity,watanabe1972densities,lohofer2004thermophysical}.

Such changes in concavity of the liquid properties, because they can be directly linked to aspects of the Gibbs energy, indicate the existence of atomistic ordering events in the liquid\cite{butler1932thermodynamics,kaptay2019improved,Tanaka1999,tanaka2006evaluation,hajra1996applicability}. Conventionally, observation of intermetallics at low temperature are often used to justify the existence of such ordering. But in the case of Cu-Ni, the liquid appears to behave as if such associates may exist without obvious appeal to ordered solid phases at low-temperature\cite{lupis1983chemical}. 

Experimental thermodynamic measurements of Cu-Ni have yielded large, positive enthalpies of mixing in the solid-solution and liquid, indicating strong non-ideality in high-temperature Cu-Ni alloys in both the solid and melt \cite{dench1963adiabatic,srikanth1989thermodynamic,Elford1969,Turchanin2007,Rapp1962}. 

Using Equation (\ref{configurationalentropy}), the feature-set and shape of simple lens- melting phase diagrams, like the one depicted in Figure (\ref{CuNiPhaseDiagram}) are computed analytically using the uncorrelated configurational entropy of mixing ($\Delta S^{\text{config}}$)\cite{Fowler}:

\begin{equation}
\begin{aligned}
    \Delta S^{\text{config}} &= \Delta S^{\text{mix}}_{\text{disordered}} \\
    & = -R\left ( x\textsubscript{Cu}\ln x\textsubscript{Cu} + x\textsubscript{Ni} \ln x\textsubscript{Ni} \right)
\end{aligned}
\label{configurationalentropy}
\end{equation}

and simple enthalpy term ($\Delta H$) of the type:

\begin{equation}
\Delta H = \Omega_{AB}x_{A}x_{B}
\label{simpleEnthalpy}
\end{equation}

Where $\Omega_{AB}$ is related to the difference in bonding energy of the homogeneous (AA,BB) and heterogeneous type (AB) bond, $x_i$ are the mole fractions of the constituents, and R the ideal gas constant.

Gibbs energy curves can be manufactured that reproduce the observed phase transition boundaries through manipulation of the enthalpy term, despite that fact that a non-zero enthalpy of mixing would imply non-random ordering and therefore an altered entropy term to account for homo and hetero-clustering of atoms.  

Previous work has sought to amend the configurational entropy by relaxing the fully-random ordering assumptions on the entropy of mixing in metallic systems, for example using cluster (or associates) models \cite{Cowley1950,Cowley1960,kikuchi1977vi,lupis1983chemical}. However, a pressing challenge associated with such models is that the choice of the associate and their number is rather arbitrary, and in many cases, is used as a fitting factor to reproduce observed phase behavior as illustrated in appendix A for Cu-Pd. This is problematic in the case of Cu-Ni as there are no low-temperature intermetallic compounds to appeal to for a guess at the ordering tendencies of the high-temperature solid and liquid. Therefore, such cluster models on their own frequently also results in Gibbs energy curves with non-physical temperature variation and entropy \footnote{The necessity of appealing to the existence of associates in systems with simple lens melting and extensive solid solutions has already been justified in the literature (for example, see Cu-Pd in Appendix A).}.

It would be of great benefit to devise a new method to identify such associates or clusters, especially when they prove difficult to observe in scattering experiments because of melt reactivity and the short-range nature of the atomic ordering. Indeed, even when such cluster models (see Appendix B) successfully capture the activity coefficient, they may still incorrectly inform the enthalpy and entropy of mixing (see Appendix D). Independent evaluation of either the enthalpy or entropy of mixing is necessary to help constrain models of the configurational ordering in their evaluation of the chemical potential, and by extension, the Gibbs energy. 

Entropy can arise from the vibrational, configurational, magnetic, and electronic degrees of freedom. Whereas the configurational and vibrational contributions to the entropy have been thoroughly explored experimentally and computationally, it is Fultz who points out that, because of the saturation in vibrational modes and the existence of an upper bound on the configurational entropy, at high temperatures ( $>$  0.7 $T_{m}$), the electronic contribution to the entropy may prove increasingly important for metal mixing thermodynamics\cite{fultz2010vibrational}. So far, such electronic contributions have been ignored in lens-type metal binaries. 

The pervasive belief that the electronic contribution to the entropy is small in metal and their alloys contrasts with insights from metal electronic structures. The electronic contribution to the entropy is quantified by low- temperature ($<$ 10K) calorimetric measurements that evaluate the heat capacity. The linear nature of the low- temperature heat capacity is conventionally attributed to the electrons and is given by the Equation (\ref{HeatCapacityElectronicEntropy}) \cite{sommerfeld1928elektronentheorie,ashcroft2022solid}: 

\begin{eqnarray}
\label{HeatCapacityElectronicEntropy}
    C_{V}^{electronic} = \gamma T
					\end{eqnarray}
where $\gamma$ is the slope of its linear variation with temperature. This value tends to be small for alkaline and alkaline-earth elements, which do not have large densities of states near the Fermi-level. But for transition metals, they can be quite large \cite{hultgren1973selected}. The $\gamma$ term can in principle be computed from the complete band structure of a material, but modern Density Functional Theory (DFT) methods are reported to struggle to provide accurate accounts of the electronic entropy and capture aspects of disorder in the atomic environment of solid-solution alloys. Electronic interactions and large density of states found in the transition (e.g. Mn, Ti, Cr) or magnetic (Fe, Ni, Co) metals, and the long length-scales of interaction are important in metallic systems in both their liquid and solid states \cite{ashcroft2022solid,kardar2007statisticalfields}.

Calculations of the entropy of mixing and the resulting phase behavior that have included such electronic contributions for select liquid transition metal systems do in fact suggest a strong role for these electronic degrees of freedom in and near the melting point \cite{khanna1980entropy,pasturel1985electronic}. 

But using low-temperature calorimetry has limitations for multi-component systems and liquids. The first assumption when extrapolating low-temperature heat capacities is that the electronic contribution only changes linearly with temperature. If there are any phase transformations, it is anticipated that the electronic heat-capacity itself would change discontinuously at the phase transformation temperature. Additionally, low temperature linearity in the heat-capacity can be shown to arise from other material phenomena, including crystal field level splitting and Kondo-type spin fluctuations\cite{manley2002vibrational,manley2003importance,paglione2003field}. Knowledge of the electronic entropy gathered independently of the heat-capacity may help in extracting the electronic contribution from other confounding effects. Otherwise, projection of the electronic heat capacity remains a qualitative in nature at higher temperatures. 

Recent work has proposed to experimentally evaluate the electronic entropy using transport properties and to rationalize these findings with the total entropy of order-disorder phase transitions in Cu$_{3}$Au and metal- insulator transitions in VO$_{2}$ \cite{paras2020electronic,paras2021contribution} as well as in liquid semiconductors \cite{Rinzler2016,rinzler2017thermodynamic}. This method has yet to be extended to metallic chemistries at high-temperature for solid solutions and liquid metal alloys. 

Cu-Ni is an exemplary system for several reasons. The conventional methods of understanding the electronic entropy indicate that Ni, owing to its suspected large density of states at the Fermi-level, should have a large electronic entropy, whereas Cu would not \cite{hultgren1973selected}. Table (\ref{CuNiCalorimetricEentropy}) reports estimates of the electronic contribution from the linear temperature variation of the electronic heat capacity across the alloy range, indicating a large contribution to the electronic entropy in much of the solid solution. The large electronic entropy indicated by calorimetric measurements and a lack of inclusion in thermodynamic modeling, suggests that perhaps the random atomic ordering in the solid and liquid alloy may not be representative of neither the spatial configurational ordering nor the thermodynamic reality.  

Transport properties have been measured in liquid Cu-Ni alloys but not as thoroughly in the high-temperature solid solution \cite{ZytveldCuni}. 
Herein we report experimental thermopower and electrical conductivity measurements for such high-temperature solid solutions. The results indicate significant electronic contributions to the entropy in solid-solution alloys and that transport property measurements at high-temperature may be useful in identifying atomic ordering in nominally "ideal" systems.  

\begin{table*}[htbp]
\centering
\begin{tabular}{ccccccccc}
\toprule
xCu (at.$\%$) & 0  & 0.184 & 0.38 & 0.579 & 0.63 & 0.784 & 0.892 & 1 \\
\midrule
$\gamma$ (J/mol K$^{2} \times 10^{4}$) & 70.29 & 66.9 & 63.6 & 69.45 & 40.17 & 19.12 & 10.46 & 6.7  \\
\bottomrule
\end{tabular}
\caption{Low temperature $\gamma$-coefficient from Equation (\ref{HeatCapacityElectronicEntropy}) for Cu-Ni alloys compiled by \cite{hultgren1973selected}.}
\label{CuNiCalorimetricEentropy}
\end{table*}

\section{Methods}

\subsection{The Electronic Entropy}

It has been proposed that the partial molar entropy of a conduction electron can be related to the thermopower from \cite{Callen1948,Rockwood1984}.
		
					\begin{eqnarray}
					\left( \frac{\text{dS}}{dn_{e}} \right)_{T,P,n_{j}} = - \alpha F
\label{partialselectron}
\end{eqnarray}
where \emph{F} is Faraday's constant and $\alpha$ is the thermoelectric power.

The integral form of this equation was derived in \cite{rinzler2017thermodynamic,Rinzler2016} and resulted in the
electronic state entropy:
					\begin{eqnarray}
					S^{\text{e}} = - n_{e}e\alpha
      \label{StateElectronicEntropy}
					\end{eqnarray}
    
where \emph{e} is the fundamental charge constant, $n_{e}$ is the number of
free charge carriers (here electrons). 

This formalism has been applied to evaluate the electronic contribution to the mixing entropy in liquid Te-Tl alloys, as well as the electronic entropy for the metal-insulator transition in solid VO\textsubscript{2} and the order-disorder transformation in solid Cu$_{3}$Au \cite{rinzler2017thermodynamic,paras2020electronic,paras2021contribution}.

Our previous work in this area focused on the contribution of electrons to first-order phase transformations\cite{paras2020electronic,paras2021contribution}. The objective of this article is to explore the temperature and composition variation of these transport properties in metal alloys. The entropy metric itself will be slightly different, instead of focusing on the total entropy of a phase transition, we will focus on the mixing entropy. Because the entropy of mixing is a function of both the entropy of the mixed state and the endmembers, Cu-Ni was chosen to avoid allotropic transitions as might occur in other alloy systems with large solid solutions like Fe-Ni and Fe-Cr. Our goal is then to study how the electronic contribution to the entropy of mixing may reveal aspects of atomic and electronic ordering that are not indicated in otherwise unremarkable phase diagrams of such solid-solution alloys. 

\subsection{Experimental Methods}
Samples of Cu-Ni alloys were prepared in-house using a Buehler AM-500 arc-melter by combining pure slugs of Cu and Ni. Total sample mass was kept consistently 50 g.  Slugs of Cu and Ni were sourced from Thermo Fischer (Puratronic, 99.99$\%$). Samples were remelted 5 times under Ar cover gas and flipped each time to improve alloying homogeneity. Between each melt, the system was purged using a diffusion pump to 10$^{-6}$ mbar and a sacrificial titanium getter was remelted and allowed to fully solidify. The sample hearth was made of Cu and actively water cooled.  Mass loss was negligible for at less than $<$ 0.3 wt.$\%$.  After alloying, the 50 g buttons were suction cast into rectangular ingots of roughly 60 mm x 20 mm x 8 mm using identical operating conditions to the alloying process. The samples were than cold-rolled to a minimum 30$\%$ thickness reduction, sealed in a quartz ampoule with a Zr-Fe-V getter, and homogenized for 24 hours at 950$^\circ$C. Appendix C reports representative X-ray diffraction (XRD) data confirming FCC long-range ordering in the alloys. 

\subsection{Transport Measurements}

An ULVAC-RIKO ZEM-3 was used to measure both the electrical resistivity and the thermopower as a function of temperature. The furnace was in high-temperature configuration, so measurements were conducted between room temperature and 1000$^\circ$C. The system was calibrated using a Constantan calibration sample (provided by ULVAC) of identical geometry to those used in this study. Care was taken to gently lap the type-R thermocouple probes using emery or SiC paper (800-1200 grit) before and after experiments to ensure quality contact during the course of measurements. Measurements were automated using ULVAC software and were conducted under He gas after He purging of the chamber. He gas was scrubbed using a Ti gettering system from OxyGon (OG-120M) achieving $<$ 50 ppb O$_{2}$ of the outlet process gas (according to the built-in oxygen sensor). 

Each sample was measured 3 times at each temperature point, and at-least two samples were used for each experimental composition which were then averaged together for a composite curve. Error associated with the ZEM-3 measurement geometry was quoted by the manufacturer to be 7$\%$ for thermopower measurements and 10$\%$ for resistivity. Deviation among samples tested never exceeded these limits, therefore the manufacturer quoted error percentages should be used when considering expected experimental error for each data-point. The error bars on the figures will otherwise be suppressed for reading clarity. 

\subsection{Cluster and Renormalization Methods}

Prigogine and Defay have codified the treatments of solution thermodynamcis with clustering and ordering in their book on chemical thermodynamics, in particular for low-temperature organic solutions \cite{prigogine1958chemical}.  The details of the model used are given in Appendix B. The methodology is akin to a renormalization of the interaction length-scale to the level of several atom clusters, called associates, with the assumption that the resulting solution is an ideal-associated solution (no interactions between associated particles and free atoms).

\subsection{Assumptions Concerning the Electronic Structure}

The existence of multiple carriers would change the interpretation of transport properties in this paper in both the solid and liquid. Equation (\ref{partialselectron}) can be revised to introduce terms associated with additional carriers in metals. In this vain, significant work has already been undertaken to try and understand the Hall effect and thermopower in metals and alloys using such an approach \cite{hurd2012hall,takano1967interpretation,hitchcock1971Hall,harman1967thermoelectric,dugdale1969Hall,tsuji1958thermoelectric}.  Despite these efforts, even theoretical evaluations of the thermopower of alkali metals, like Li, exhibit neither the right slope nor the correct sign \cite{xu2014first}. 

To integrate the experimental results with conventional understanding of multi-carrier effects on the electronic transport properties, we will assume the following:

\begin{itemize}
    \item The electronic structure of a material can be parameterized in terms of several electron bands
    \item There will be no consideration of scattering between bands and each will be treated as a separate electronic conductor
    \item The electronic entropy derived from calorimetric measurements (Equation (\ref{HeatCapacityElectronicEntropy})) is equal to the transport electronic entropy at low-temperatures 
    \item Defining thermopower and conductivity contributions discretely indicates we are not considering the effects of correlation among either like or unlike type charge carriers
\end{itemize}

 The first point is taken out of necessity; while band theory has managed to explain mechanical and electronic behavior in some simple metals, the solution to an electron wave equation proves intractable under the assumptions necessary to ensure the physical accuracy for solution thermodynamic work. We lean on these concepts not out of endorsement of the approach but in lieu of other methods to describe metallic chemical bonds. The second point is a gross oversimplification, and may render our conclusions qualitative in nature. Electron scattering simulation remains an open frontier in condensed matter physics.  The last point can even be considered a hypothesis in its own right, as agreement between the transport method for deriving an electronic entropy, and the method of calorimetry, have not yet been shown to describe the same quantity \cite{ashcroft2022solid}. We will however make this assumption and examine its consequences.

\subsection{Derivation of the Effect of Multiple Carriers on the Electronic Entropy}

Because of the form of the state electronic entropy in Equation (\ref{partialselectron}), it is more practical to assume that the carrier type does not change across the phase diagram. We will explore the validity of this assumption in the context of the Cu-Ni system and demonstrate that while this may not be fully true on the Cu-rich side of the phase diagram, for this particular alloy system, deviations from single-carrier conduction due to complicated Fermi-surface geometry is not quantitatively important for understanding the systems mixing thermodynamics. We will first take the approach of Hitchcock and Stringer, whose own work extended Takano and Ziman's\cite{ziman1961ordinary,takano1967interpretation,hitchcock1971Hall}.

The low field Hall effect when considering two bands, one of holes and one of electrons is given by
    \begin{equation}\label{two-band Hall}
    R_{H} = \frac{1}{\sigma^{2}e}\left( \frac{\sigma_{h}^{2}}{n_{h}}-\frac{\sigma_{e}^{2}}{n_{e}}    \right) 
    \end{equation}
where $\sigma$ is the total conductivity, $n_{e,h}$ are the quantities of itinerant electrons and holes, $\sigma_{e,h}$ are their relative contributions to the conductivity, and $\sigma$ is given by
    \begin{equation}\label{totalConductivity}
    \sigma=\sigma_{e}+\sigma_{h}
    \end{equation}
The conductivity can also be recast in terms of kinetic properties of those carriers as
    \begin{equation}
    \sigma_{i} = \frac{n_{i}e_{i}^{2}\tau_{i}}{m_{i}} 
    \label{ballisticConductivity}
    \end{equation}
where $e$ is the fundamental charge, $\tau_{i}$ is the relaxation time and is associated with the various collision processes that electrons and holes undergo with applied electric field, and $m_{i}$ is the effective mass of the carrier. Because the electrical conductivity of a band can be written as a function of energy, $\sigma_{E}$, more complicated implementations of transport models which include non-constant relaxation assumptions can be incorporated  in Equation (\ref{ballisticConductivity})\cite{Xu2014}. 

Under the one-band interpretation of the Hall effect, one assigns an "effective" number of one-band electrons, which we denote $n^{*}$. Then, using the two-band model the expression for $n^{*}$ in terms of the various parameters becomes 
\begin{equation}\label{nstar}
				n^{*} = \frac{n_{e}n_{h}(1+\sigma_{h}/\sigma_{e})^{2}}{n_{h}-n_{e}(\sigma_{h}/\sigma_{e})^{2}}
\end{equation}
Combining Equations (\ref{ballisticConductivity}) and (\ref{nstar}), the ratio of conductivities is:
\begin{equation}\label{conductivityRatio}
				\frac{\sigma_{h}}{\sigma_{e}}=\frac{n_{h}m^{*}_{e}\tau_{h}}{n_{e}m^{*}_{h}\tau_{e}}
\end{equation}
From Equations (\ref{two-band Hall}),(\ref{nstar}) $\&$ (\ref{conductivityRatio}) the number of electrons and holes and their interpretation of the Hall-coefficient will require knowing the ratio of the effective masses, the total Hall effect which is experimentally accessible, and the ratio of their relaxation times. Complicated anisotropy and large effects on the energy dependence of these properties can be included through precise evaluation of these various electronic structure properties. However, for subsequent evaluations, we will assume identical relaxation times for the electrons and holes, but provision for differences in their effective masses and quantities \cite{takano1967interpretation}.

The electronic entropy of a multiband model is given by
				\begin{equation}
				\alpha = \frac{\sigma_{e}\alpha_{e}+\sigma_{h}\alpha_{h}}{\sigma_{e}+\sigma_{h}} 
        \label{two-band thermopower}
				\end{equation}
The same parameters are necessary to evaluate the relative contribution to the partial molar electronic entropy of each carrier. The final revision to the electronic entropy equation under these simple assumptions is given  by Equation (\ref{MultibandEntropy}).

					\begin{eqnarray}
					\label{MultibandEntropy}
					\ S^{e}=-n_{n}e\alpha_{n}+n_{p}e\alpha_{p}
					\end{eqnarray}

Where $n_{i}$ and $\alpha_{i}$ are the number of carriers and the contribution of those carriers to the thermopower in the electron (n) and hole (p) subands respectively. These result suggests that changes in the sign of the thermopower and Hall-effect may not only offer the possibility of a negative partial molar electronic entropy as discussed in \cite{paras2021contribution}, but that there may exist multiple carriers with distinct contributions to the electronic entropy which are masked by their offsetting contributions to the thermopower. Examining Equation (\ref{two-band thermopower}), metals that have large thermopowers have electronic entropies that should be dominated by a single band: any additional carrier types would necessarily reduce the thermopower, which is typically small for metals in general. Small magnitudes of the thermopower may occur in metals either because their is compensation between multiple carrier types  or a minimum in the electronic entropy.

\section{Results}

\subsection{Transport Properties}
\begin{figure*}[!htb!]
		\includegraphics[width=\linewidth]{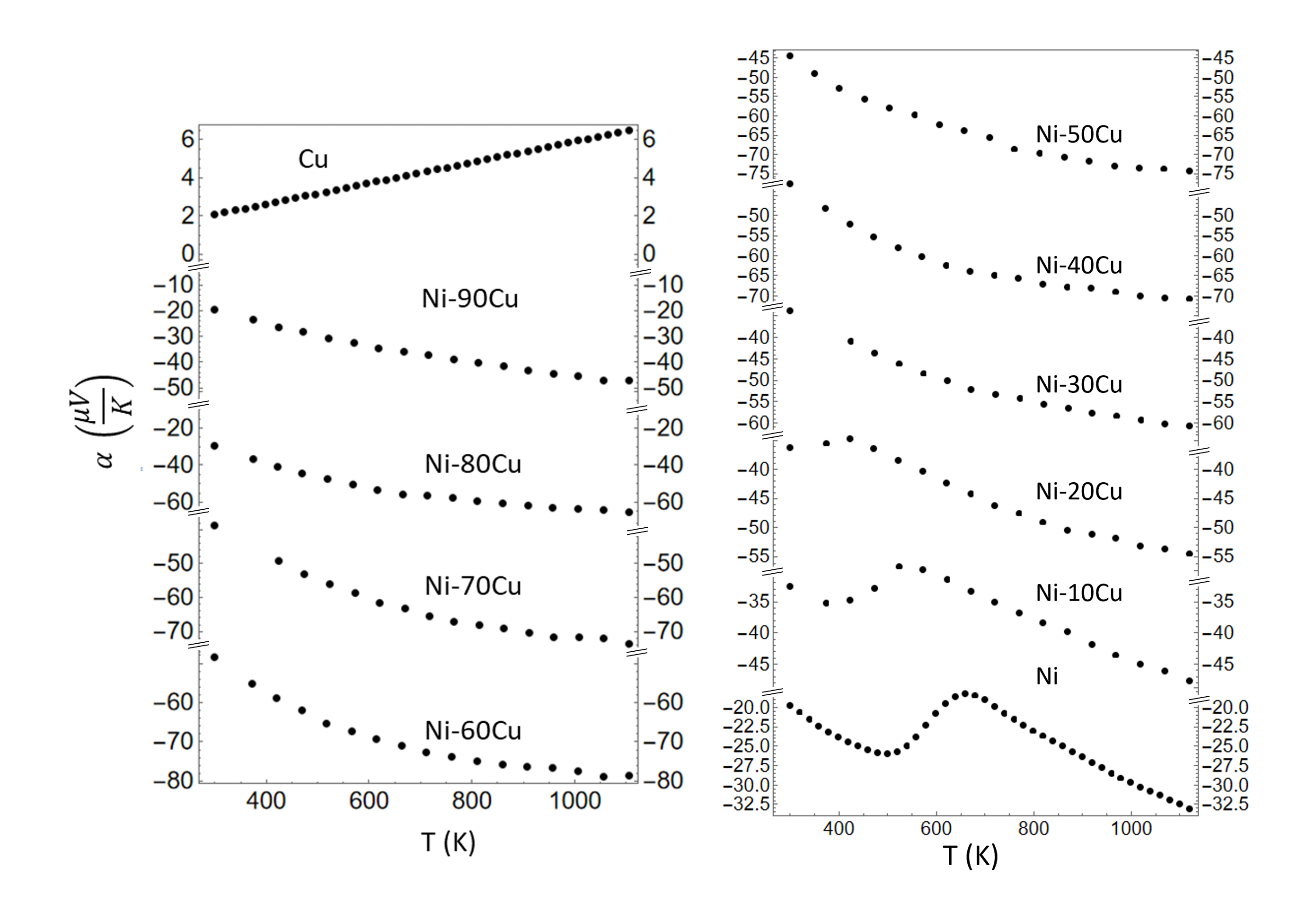}
				\caption{Absolute thermopower ($\alpha$) were measured for the Cu-Ni alloy system as a function of temperature. The temperature and composition dependence are plotted here. Note the discontinuous y-axis.}
    \label{CuNiThermopower}
				\end{figure*}

\begin{figure*}[htb!]
		\includegraphics[width=\linewidth]{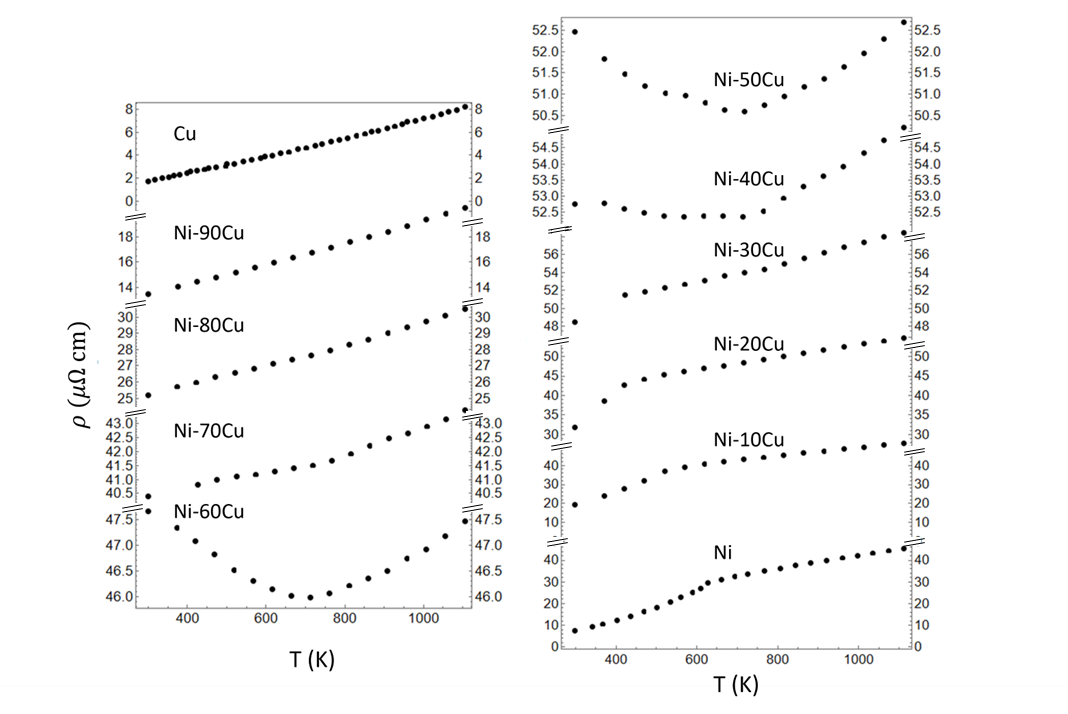}
				\caption{The resistivity were measured in the Cu-Ni alloy system as a function of temperature. The temperature and composition dependence are plotted here. Note the discontinuous y-axis.}
    \label{CuNiResistivity}
				\end{figure*}

The absolute thermopower and the resistivity were measured for 9 different Cu-Ni alloys, with supporting data for the end-members from \cite{Abadlia2014}. They are plotted in Figures (\ref{CuNiThermopower}) and (\ref{CuNiResistivity}) respectively. The thermopower of Cu-Ni alloys monotonically increases in magnitude with temperature except for Ni rich samples, which go through both local minimum and maximum as they approach their Curie temperature. One can observe from both figures the monotonic reduction in the Curie temperature shown in Figure (\ref{CuNiPhaseDiagram}). An anomalous minimum in the resistivity emerges from 70-30at.$\%$Cu which was also observed by others\cite{Ahmad1974}.  These values compare well with prior measurements of thermopower and resistivity measurements at high-temperature from Ahmad and Greig \cite{Ahmad1974}.

\subsection{Electronic State Entropy of Pure Ni}

\begin{figure}[h]
				\includegraphics[width=0.5\textwidth]{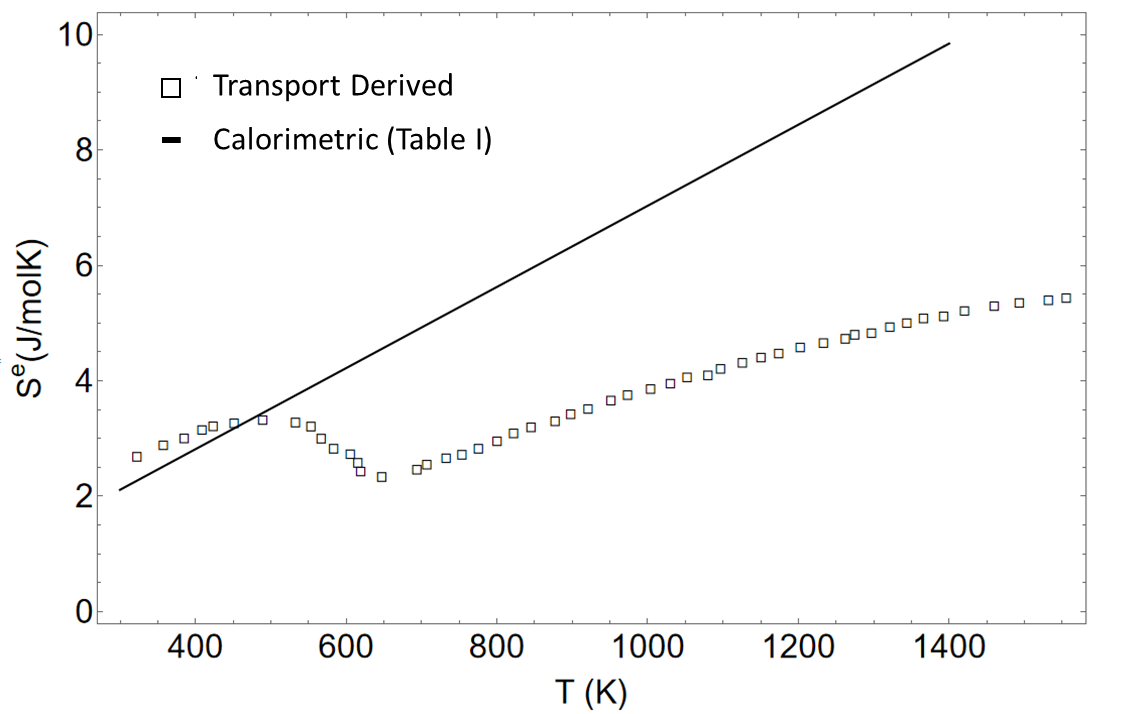}
				\caption{The calorimetric and one-band transport electronic entropies of pure Ni as a function of temperature. The deviation between the two curves due to the effect of demagnetization of Ni at 627 K increases as a function of temperature.}
				\label{pureNielectronicentropy}
				\end{figure}

The thermopower and Hall effect are both negative for pure Ni \cite{Abadlia2014,hurd2012hall}. Furthermore, the thermopower is large in magnitude for a metal ($-20 \mu$V/K) whereas typical values are O(1). Low-temperature calorimetric measurements of the electronic component of the heat capacity compiled by Hultgren and calculated using Table (\ref{CuNiCalorimetricEentropy}) and Equation (\ref{HeatCapacityElectronicEntropy}) suggest that the electronic entropy of Ni is also quite large even at room temperature \cite{hultgren1973selected}. Using the single band model in Equation (\ref{partialselectron}) and Hall effect data from \cite{pugh1955band,perez2020entropy}, the calorimetric and one-band transport electronic entropies are compared in Figure \ref{pureNielectronicentropy}.

The pure Ni electronic entropy begins to deviate from the calorimetrically indicated value as Ni approaches its Curie temperature and the impact of demagnetization on the electronic transport properties are observed. Because the calorimetric method relies on low-temperature measurement, comparison above the curie temperature is not fruitful beyond order-of-magnitude. Pugh and Rostoker conducted ordinary Hall effect measurements on Fe and Ni\cite{pugh1953Hall}. They found that while there were holes present in pure Ni the number of carriers indicated by one-band transport analysis did not differ by more than a factor of 2 than with the multi-band model implemented herein. When allowing for two carrier types, it was found Ni exhibited 0.6 electronic (s-like) carriers per atom, and 0.6 hole (hole-like) carriers\cite{pugh1953Hall}. They also found that these results were only plausible with a ratio of the hole-like carrier conductivity to the total conductivity as $\frac{\sigma_{h}}{\sigma}=0.23$. Evaluating Equations (\ref{two-band thermopower}) and (\ref{MultibandEntropy}) using the room-temperature calorimetric entropy as a constraint, the sub-band thermopower coefficients can be calculated. This results in an $\alpha_{e}=-28$ and $\alpha_{h}=8.5 \mu V/K$; most of the electronic entropy in Ni is located in the electron subband.

\subsection{Electronic State Entropy of Pure Cu}

\begin{figure*}[ht]
    \centering
    \hspace*{-0.1\linewidth} 
    \includegraphics[width=1.2\linewidth]{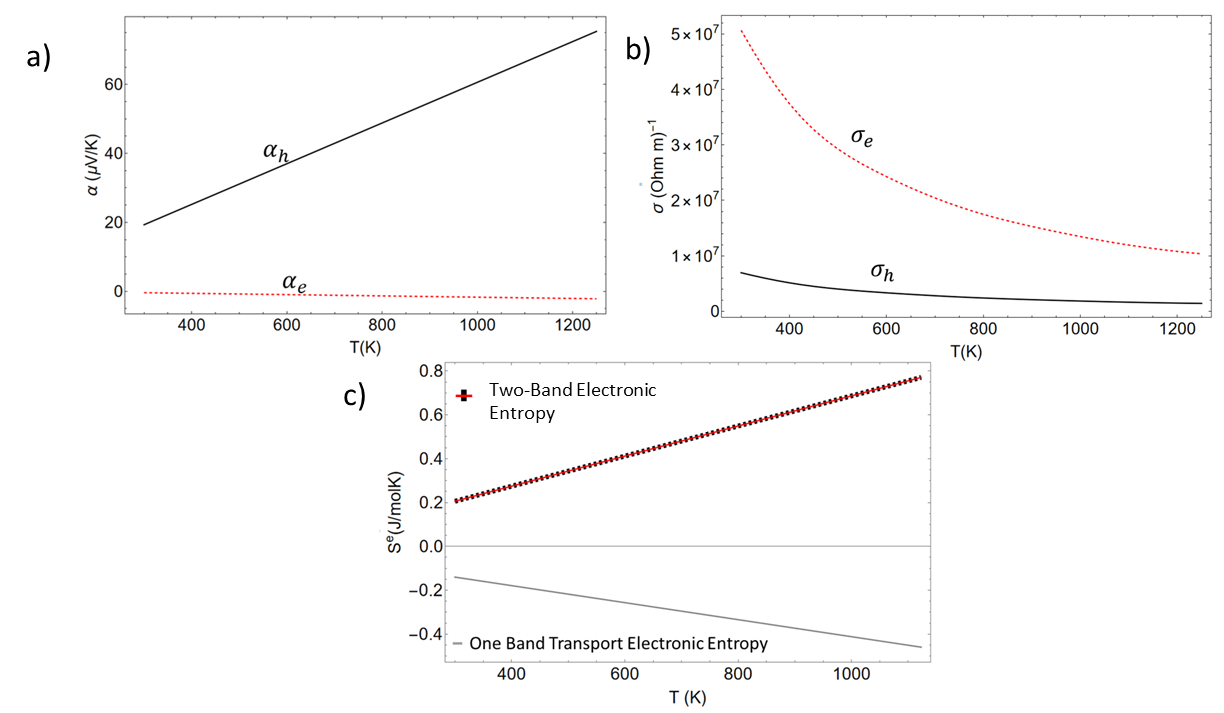}
    \caption{a) The electron belly and hole neck subband contributions to the total thermopower in pure Cu. b) The relative apportionment of the hole and electron contribution to the conductivity of those subbands. c) The evaluation of the electronic entropy in a one-band and two-band model for pure Cu metal}
    \label{PureCuPicture}
\end{figure*}

Pure Cu exhibits a positive thermopower over the entire temperature range of interest (room temperature through melting) and a negative Hall effect coefficient \cite{hurd2012hall,Abadlia2014}. Significant work has been conducted to try and understand why this is the case, with various models proposed by Ziman, Hitchock, Singer, and Takano \cite{takano1967interpretation,hitchcock1971Hall,ziman1961ordinary}. The overarching theme has been that there exists some protuberance in the Fermi surface of copper, typically along the $<111>$ family of directions in reciprocal space, and the neck-regions that meet at the first Brillouin zone produce regions of negative curvature, resulting in hole-like states that are responsible for the positive thermopower. Many of the studies cited have conducted parametric studies of the parameters in Equations (\ref{two-band Hall}) and (\ref{two-band thermopower}) to determine whether upon equating the calorimetric and transport methods, one could determine a physical set of parameters for the electronic and hole contribution to the thermopower and resistivity. The results of this are plotted in Figure (\ref{PureCuPicture}), which demonstrates that the electronic entropy of Cu, if implemented in the one-band model, results in a negative electronic entropy (Figure \ref{PureCuPicture}-c) with respect to what has been measured calorimetrically at low temperature. Using Ziman's model for the number of neck vs. belly electrons derived from de Haas van Alphen measurements, we find that there a small number of holes (n$_{h}$ = 0.093 per atom) with a large subband thermopower ($\alpha_{h} = 20 \mu V/K$) with a large number of electrons (n$_{e}$=0.907 per atom) and a small subband thermopower ($\alpha_{e} = -0.397 \mu V/K$).

This result differ from those of Aldersen, Farrel, and Hurd, who indicated very different distribution of the number of carriers and assumptions around the effective mass of these materials  \cite{alderson1968Hall}. Incorporating their values into our calculations results in subband entropies that were positive for the electron subband and negative for the hole, which in effect implies a negative electronic entropy.

\subsection{Electronic State Entropy of Cu-Ni Alloy}

\begin{figure}[h]
                \includegraphics[width=\linewidth]{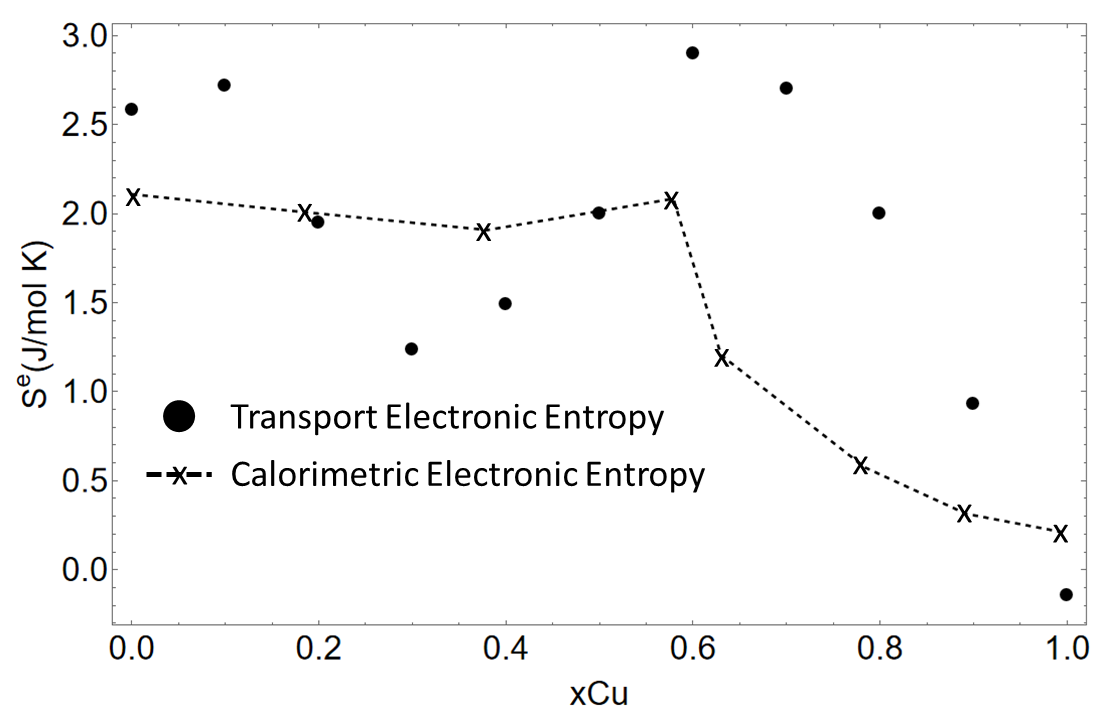}
				\caption{The electronic state entropy calculated from both transport properties and the measurements in Table (\ref{CuNiCalorimetricEentropy}) at room temperature (300 K)  .}
				\label{CuNiMultipleEntropy}
				\end{figure}

A plot of the two methods of evaluating the electronic entropy for Cu-Ni alloys is depicted in Figure (\ref{CuNiMultipleEntropy}), where the broad trends agree. This result substantiates our justification for the use of a one-band model and indicates that there may be significant electronic ordering as a function of temperature that the low-temperature calorimetric methods do not capture.  

\subsection{Unipolar Electronic Entropy of Mixing in Cu-Ni}

\begin{figure}[h]
				\includegraphics[width=\linewidth]{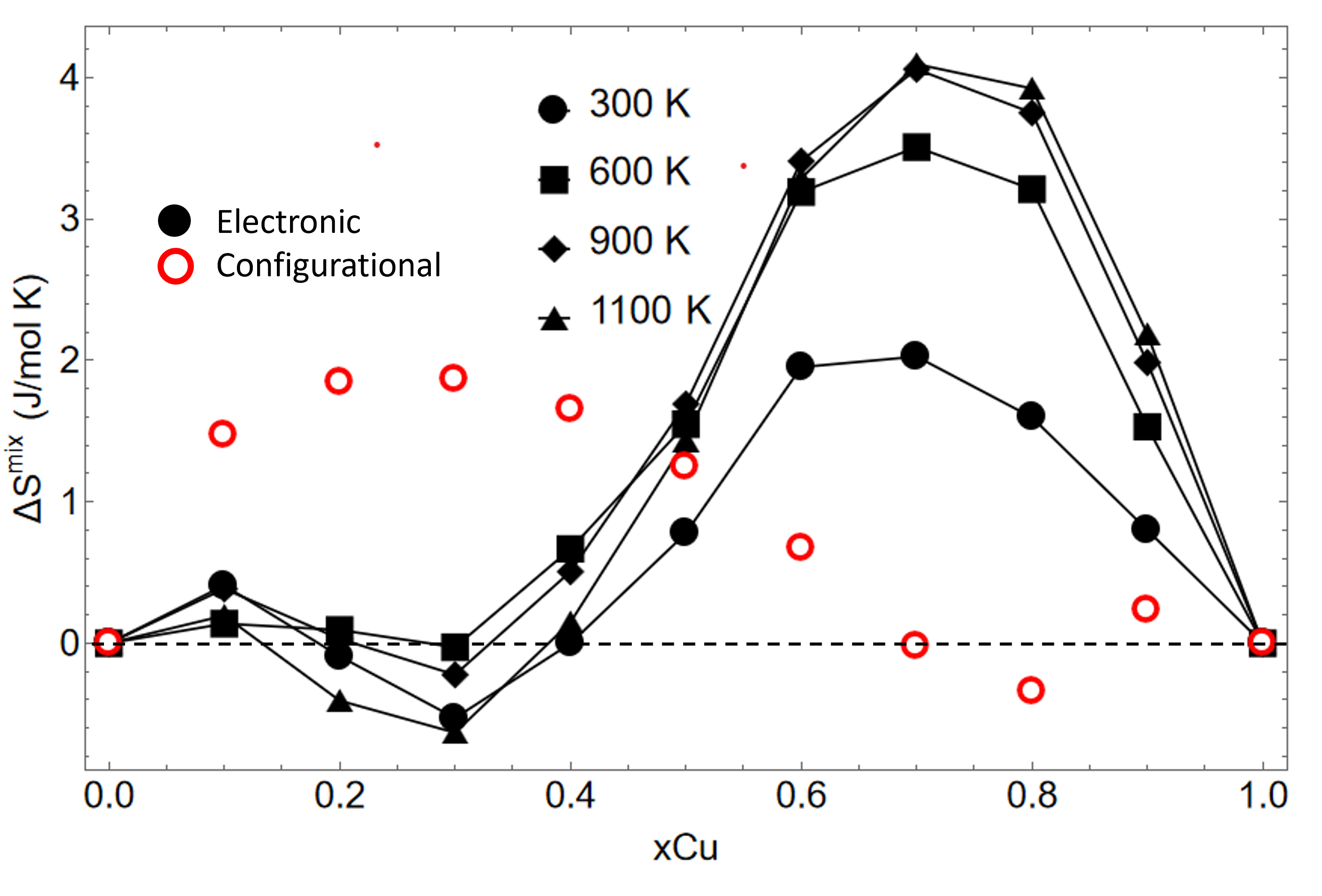}
				\caption{The electronic entropy of mixing (solid) calculated using Hall effect measurements from \cite{perez2020entropy} and the cluster configurational entropy of mixing using the method of Shunyaev and Lisin \cite{shunyaev2002arbitrary} (empty).}
				\label{CuNiEEntroypMixing}
				\end{figure}

The electronic entropy of mixing can be defined as the difference between the composition weight partial molar electronic entropy of the end-members as

				\begin{equation}
				\Delta S^{e}_{mix} = S^{e}_{A_{x_{A}}B_{1-x_{A}}} -x_{A}S^{e}_{A}-(1-x_{A})S^{e}_{B}	
    		\label{ElectronicEntropyofMixing}
\end{equation}
Where $S^{e}$ has the definition in Equation (\ref{StateElectronicEntropy}). The electronic entropy of mixing for the Cu-Ni binary is plotted as a function of composition and temperature in Figure (\ref{CuNiEEntroypMixing}). Making use of Hall effect data from Perez et al, our own electrical transpot measurements, and Equation (\ref{ElectronicEntropyofMixing}), we find that the electronic entropy of mixing can be evaluated for this system. 

To the best knowledge of the authors, this is the largest electronic entropy of mixing reported for a binary alloy via methods other than calorimetry. The mixing entropy is large and positive on the Cu-rich side of the phase diagram, with a broad maximum near 70-75at.$\%$ Cu. There is a local minimum on the Ni rich side at 30at.$\%$Cu. As the temperature is increased, both the minimum and maximum become more profound. It appears that near 900 K, the temperature variation of the electronic entropy begins tends toward zero in conjunction with the temperature variation of the thermopower.

\subsection{Cluster Model Results}

A cluster model was implemented that made two assumptions. First, due to the broad positive enthalpy of mixing reported in Figures (\ref{EnthalpyMixCuNiSolid}, \ref{EnthalpyMixCuNiLiquid}), it was assumed that self-associates of the end-members could form in the melt and high-temperature solid-solution. The self-association energies were calculated based on a method for evaluating the entropy of fusion discussed by Shunyaev and Lisin \cite{shunyaev2002arbitrary}. 

It is clear from Appendix D that the presumption of self-association can be made to fit the activity for the system, at the cost of a quantitatively accurate enthalpy and entropy of mixing. The maximum of the Cu-Ni electronic entropy of mixing is both large and constant with temperature around 75$\%$ Cu. We therefore justify the choice of a Cu$_{3}$Ni associate on the strong mixing tendency encouraged by the electronic subsystem rather than by arbitrary choice as in the Cu-Pd system (see Appendix A). The details of the cluster model implemented in this paper are listed in Appendix B.  The only fitting parameter that remains unconstrained in our solution model is the binding energy of that associate. The energy used in these calculations is given as -45 kJ/mol as it is a similar order of magnitude used for hetero-cluster binding energy work reported elsewhere \cite{prigogine1958chemical,adhikari2011disorder,shunyaev2002arbitrary}. 

The resulting configurational entropy of mixing is plotted in Figure (\ref{CuNiEEntroypMixing}). 
Assuming like-type atom mixing and the existence of an associate identified from the variation of the electronic entropy with composition, the configurational entropy is found to be complementary in shape to the measured electronic entropy of mixing, with a peak near the local minimum of the electronic entropy of mixing and a small, negative region on the Cu rich side near 90at.$\%$Cu. Adding both terms together would qualitatively produce an entropy of mixing that is similar in concavity to what we expect from an ideal solution, although a small minimum persists near the 50at.$\%$ Cu composition. 

\vspace{1cm}

\subsection{Properties of the Liquid} 

If the ordering proposed in the previous section, based on the strong electronic entropy, causes the formation of an associate that retains its local stoichiometry in the melt, there should be evidence for such an associate in the viscometric and liquid density of the alloy. Cu-Ni is one of few systems whereby the visocsity, density, and thermopower of the liquid have all been measured thoroughly across the composition range, allowing estimation of the electronic entropy of mixing of the liquid \cite{ZytveldCuni}. 
We are missing Hall-effect measurements in liquid Cu-Ni that we have for the solid. However, because the thermopower does not significantly change significantly between our measured high-temperature solid-solution and Zytvelds's liquid data, we assume the number and type of carriers does not change either\cite{ZytveldCuni}. Our expectation is that any large redistribution of electronic states, without concomitant adjustment in the other forms of the entropy, should result in a highly non-ideal solidus and liquidus curve, which is not the case for this system. In this instance, we made the assumption that there is no large redistribution in the electronic states upon melting in this alloy.

\begin{figure}[h]
	\includegraphics[width=\linewidth]{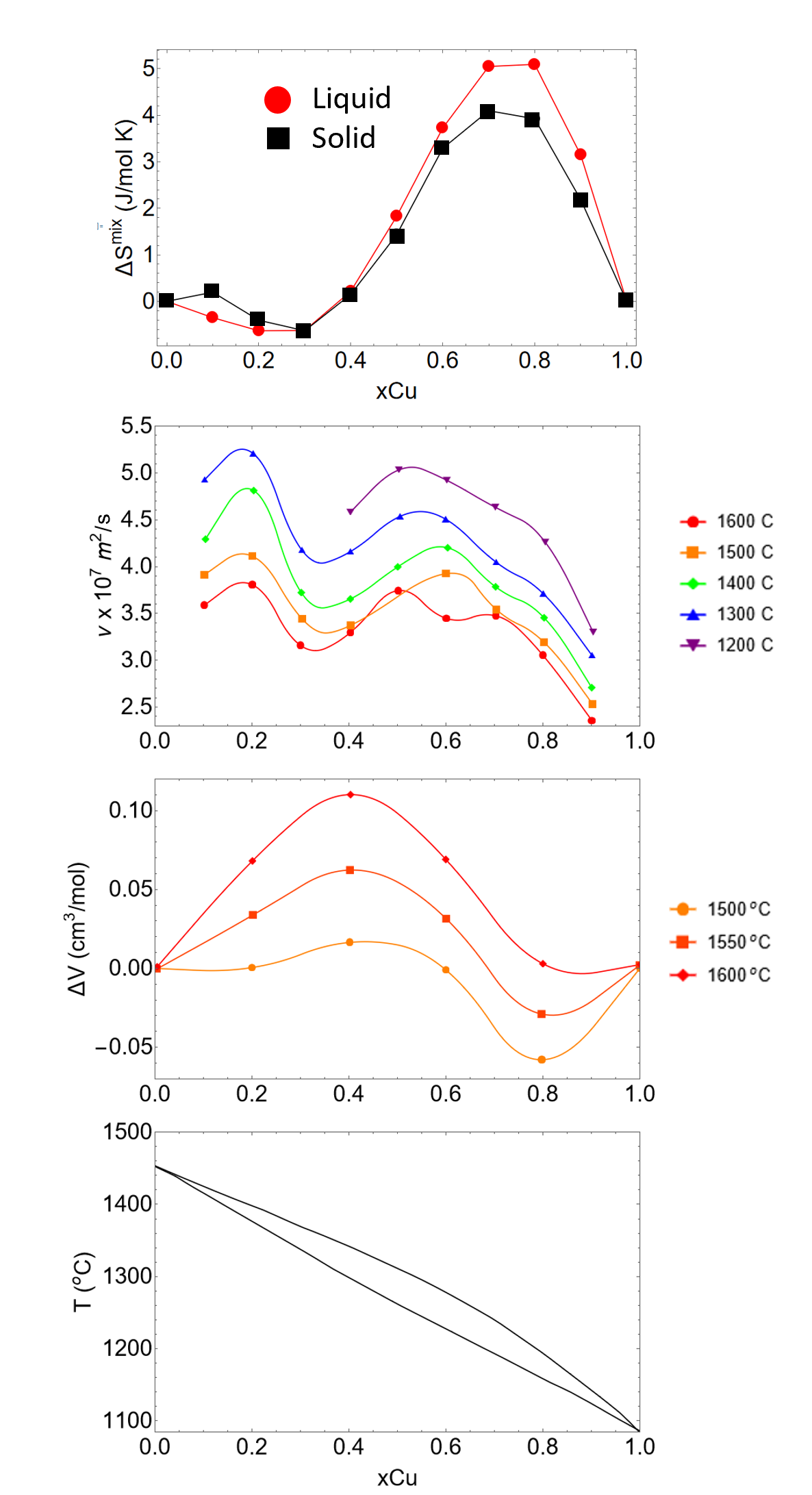}
				\caption{The electronic entropy of mixing (top panel) was computed for liquid Cu-Ni using transport properties from Zytveld and compared to the viscosity (symbol $\nu$ )and molar volume of mixing ($\Delta V$) data from Chikova and Watanabe\cite{watanabe1972densities,chikova2019viscosity,ZytveldCuni}.The bottom panel is the Cu-Ni liquidus and solidus.}
    \label{CuNiLiquidProperties}
				\end{figure}

Despite this obvious limitation, we compare the physico-chemical properties of Cu-Ni liquids and the high-temperature solid and molten electronic entropies of mixing in Figure (\ref{CuNiLiquidProperties}). The results of this comparison suggest that the type of ordering we have pursued to try and rationalize the electronic and configurational entropies of mixing for the solid-solution remain in the liquid. 

One can readily see that local maxima and minima in both the density and viscosity correspond to the compositional variation of the electronic entropy of mixing in both the high-temperature solid solution and the melt. This may prove fruitful to pursue purely empirical correlations between these quantities, as the measurement of the thermopower and resistivity are found to typically be easier than the viscosity. Similar trends between electrical and melt transport properties have been identified elsewhere in liquid Al-Cu alloys \cite{stinn2018thermodynamic}.

\section{Discussion}

\subsection{State Electronic Entropy}

To the best knowledge of the authors, this is the first time systematic comparison of the electronic contribution to the entropy from calorimetric and transport measurements have been made in an alloy and their pure endmembers.

In the case of pure Ni, we can categorize its large thermopower and large density of states as being dominated by a single carrier type. This is likely why below the Curie temperature, where the low-temperature electronic structure has yet to deviate heavily from its room-temperature counterpart, the quantitative difference between the transport and calorimetric electronic entropy is small.

Under the multi-band formalism, Cu exhibits a positive thermopower because the effective mass of the holes is smaller than the electrons in the belly region, owing to the higher degree of curvature in those regions. This masks the electronic entropy contribution if one were to use a one-band model.  More entropy is therefore transported by the holes owing to their more "gaseous" behavior (lighter mass and low density). But because of Cu's small entropy relative to the state electronic entropy of the alloy and of Ni, this difference, even if computed incorrectly, is negligible for the conclusions of this paper. 

It is worth noting, however, that if our method, anchored in thermodynamics, is found to be predictive, it is useful to constrain the many methods of electronic structure calculation and the nature and type of carriers that alloys and elements exhibit.

But even if single-band assumptions are made, the error in the electronic entropy of mixing for the rest of the system compared to what is expected from calorimetry, is small.  In the Cu-Ni alloy system, the thermopower and Hall-effect agree everywhere else. Such overall discrepancies are not critical when trying to evaluate the electronic entropy of mixing, which is more heavily dominated in this case by new states that emerge in the alloy and by the pure Ni endmember. 

While both methods of entropy evaluation share similar trends with composition per Figure (\ref{CuNiMultipleEntropy}), there is some difference expected not just because of inaccuracies in the transport interpretation but also in the thermodynamic differences between the two methods of evaluation.

Equation (\ref{HeatCapacityElectronicEntropy}) is derived from a constant volume heat-capacity. It should become increasingly inaccurate with increasing temperature. Whereas the electronic entropy from transport measurements are made at constant temperature and pressure, therefore resulting in a constant pressure heat capacity. The classic relationship between these two entities is given in Equation (\ref{cvminuscp}). This relationship may be useful if one is able to isole the electronic contribution to the thermal expansion and isothermal compressiblity for systems that do not experience phase transitions along the temperature range that the calorimetric methods are valid. But for now, if the proposed methods are to be believed, we can directly evaluate the ration between $\alpha$ and $\beta$ for the electron system.

					\begin{eqnarray}
					\label{CvTotal}
					C_{P}-C_{V}=\frac{TV\alpha^{2}}{\beta}
     \label{cvminuscp}
					\end{eqnarray}

\subsection{Entropy of Mixing}

While the combination of the electronic entropy of mixing measured in this work, and the cluster configurational entropy of mixing does reproduce the type of concave entropy of mixing one would expect from an otherwise ideal system, we have not attained quantitative agreement with the types of thermodynamic measurements and assessments summarized in \cite{srikanth1989thermodynamic}. There are several reasons for this. 

First, to keep the cluster model analytical and tractable, we did not explore the topological nature of the ordering in this system. We only assumed that the number of bonds in a cluster extended as if they were linear chains, and that such additional bonds did not have any kind of embedding term; there is no change to the binding strength of a cluster as the number of atoms change. Work by Lupis has sought to address this very point in implementation of an embedding term in the central atom model\cite{lupis1983chemical}. 

Second, the nature of the thermodynamic measurements for the activity coefficient in metallurgical systems like Cu-Ni have proven problematic. The relationship between the activity coefficient and the excess enthalpy and entropy are given by \cite{lupis1983chemical} :

					\begin{eqnarray}
					\label{lngamma}
				ln(\gamma_{i}) = H_{i}^{xs}-TS_{i}^{xs}
					\end{eqnarray}

Where $H_{i}^{xs}$ is the partial molar excess enthalpy of component $i$ of the solution and $S_{i}^{xs}$ is the excess partial molar entropy of the same. It is clear from Equation (\ref{lngamma}) that the apportionment of an activity measurement at a fixed temperature between the enthalpic and entropic contribution can only be done either with a solution model or the measurement of the temperature variation of the activity coefficient; such a measurement was only conducted by one source in the high-temperature solid solution \cite{Rapp1962} whose experiments were 300 K apart. Otherwise, an ideal entropy of mixing formalism is typically used which implies that the total integral entropy of mixing is given by Equation (\ref{configurationalentropy}). 

This has the effect of reducing all the $S_{i}^{xs}$ to zero, shoving all of the uncertainty of the non-ideal solution behavior into the enthalpy term. As pointed out by Kubachewski, this results in large disagreement between electromotive force, differential scanning calorimetry, and combustion calorimetry measurements of the thermodynamic properties of metallic solutions\cite{Elford1969}, therefore the Gibbs energy, and by extension, entropies of mixing for metallic solutions are often largely driven by the manufacture of Gibbs energy curves capable of reproducing phase boundaries or are fit to equations that have limited physical meaning, let alone having proper temperature dependence\cite{AnMey1992}. 

Even when others have demonstrated contributions to the entropy of solutions that can be large and are neither configurational nor electronic, these often go unincorporated in such thermodynamic evaluations \cite{fultz2010vibrational}.
The ramifications in this gap in the knowledge of the entropy has resulted in features of phase diagrams to appear as a consequence of these fitted Gibbs energy curves rather than by experimental observation. The existence of the low-temperature miscibility gap in Cu-Ni (see Figure (\ref{CuNiPhaseDiagram})) has never been experimentally observed in bulk samples, but is rather reported as a function of high-temperature measurements of a positive enthalpy of mixing with an often ideal entropy model \cite{Elford1969}.

It should also be clear that more high-temperature Hall-effect measurements are necessary to quantify the carrier number in the liquids of interest. In this article, we assumed that the number of carriers did not change significantly because the high-temperature thermopowers of the solid-solution and the liquid metal were largely similar. While this is a limitation of the current study, the result of a similar electronic entropy of mixing in the melt and solid is unsurprising given the seemingly ideal nature of the phase diagram. Such large redistribution would necessarily result in melt-behavior different from the lens. 

Finally, it is remarkable that the thermodynamics and ordering behavior seems to apply both to the solid-solution and the liquid. We speculate that perhaps there exist temperatures at which the correlation or mutual information between the electronic subsystem and the long-range crystalline order become disconnected, in which case, from the electronic perspective, there is no difference between the solid and the liquid despite the loss of long-range order in the latter. The use of such clusters has long been known to provide the right functional form of the Gibbs energies in solids but has never really been justified given that these systems are fully ergodic at most temperatures of evaluation, unlike their liquid counterparts \cite{kikuchi1977vi}.

\section{Conclusions} 

The electronic entropy was evaluated for the the solid-solution and liquid of Cu-Ni alloys. A cluster model was developed based on the suggestion of weak-association on the Cu-rich side. This qualitatively rationalized our reported electronic entropy contribution with our expectation for a concave entropy of mixing. Evidence of ordering in the melt was also substantiated by appeal to the density and kinematic viscosity and their compositional variation. The potential role of more complicated Fermi-surfaces on the interpretation of the electronic entropy was discussed but not found to be important for this particular alloy system.

\section{Acknowledgements}
This material is based upon work supported by AFOSR under award number FA9550-20-1-0163. We would like to thank Professor David Clarke for the use of his ZEM-3.
\section{References}
\bibliography{main.bib}

\appendix

\section{Cu-Pd Phase Diagram}
\begin{figure}[h]
	\includegraphics[width=\linewidth]{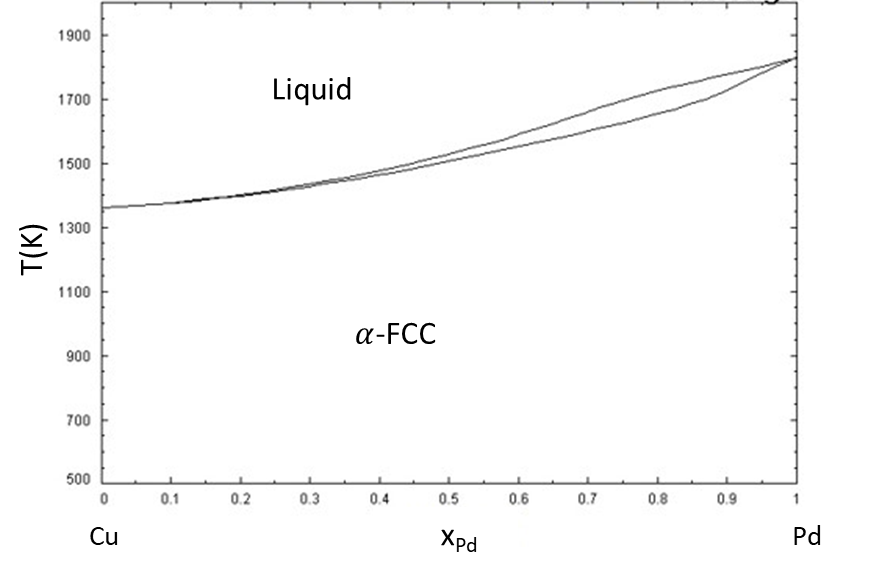}
				\caption{Cu-Pd phase diagram computed with FactSage 8.0 using the SGTE Database \cite{sundman1985thermo}, suggesting full miscibility in both the solid ($\alpha$ FCC) and the liquid state.}
    \label{CuPdPhaseDiagram}
				\end{figure}

The Cu-Pd phase diagram is quite similar to that for Cu-Ni and is plotted in Figure (\ref{CuPdPhaseDiagram}). 

This system exhibits similar features to those found in Cu-Ni. Implementing an associate of the Cu$_{3}$Pd in the liquid was found to be necessary to rationalize the modelled activity with experimental measurements and the observed phase diagram. The effect of such thermodynamic ordering in the liquid was also assumed to persist into the solid-phase of the system \cite{adhikari2011disorder}. The number of fitting parameters introduced, the nature of the interaction between those atoms in the liquid and solid and the assumption of associates of one particular stoichiometry over another remains rather phenomenological, however. 
\section{Cluster Model}
\begin{figure}[h]
        
                \centering
				\includegraphics[width=.81\linewidth]{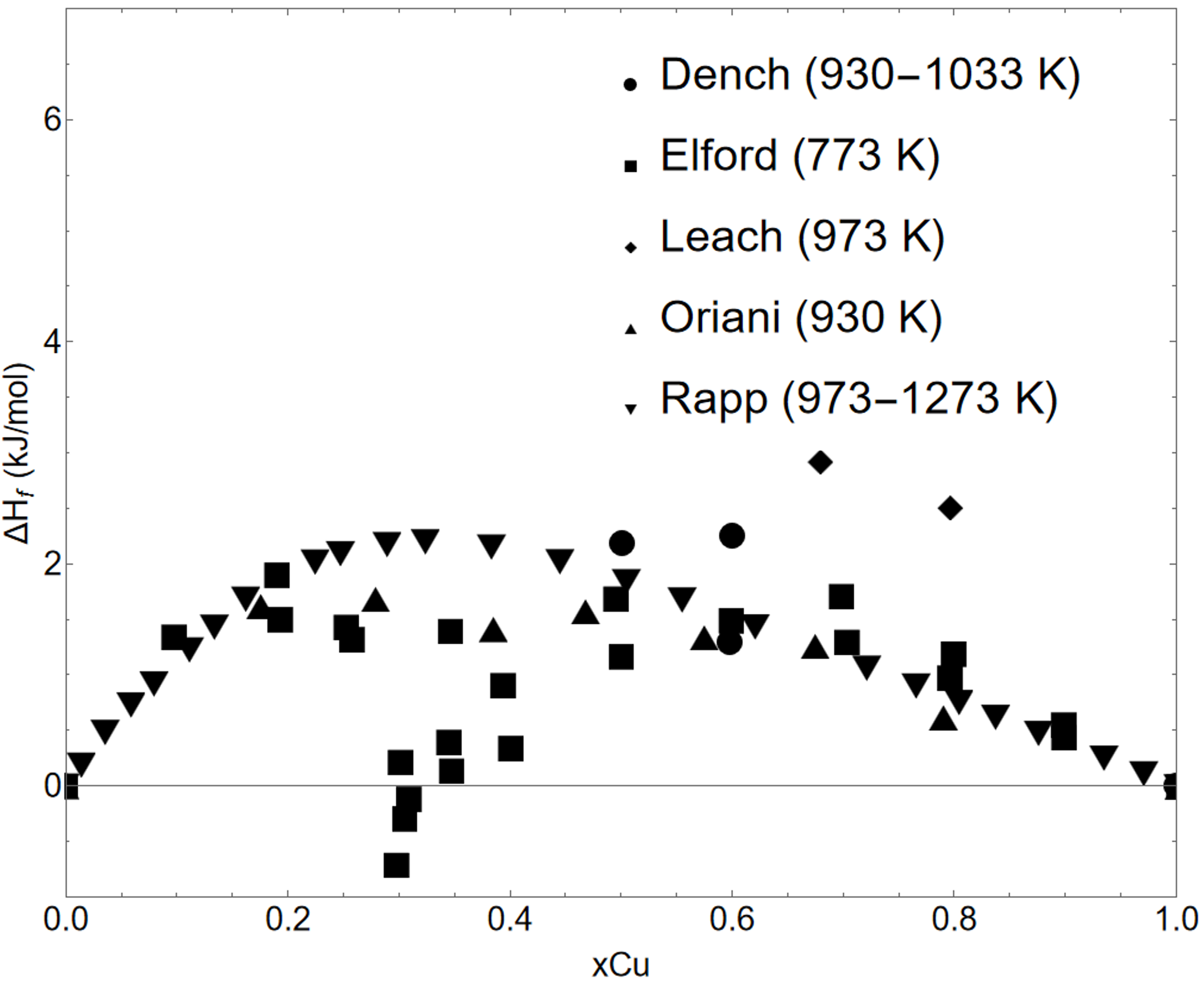}
				\caption{Enthalpy of mixing measured through calorimetric and EMF methods for solid Cu-Ni alloys\cite{dench1963adiabatic,Elford1969,Rapp1962}.}
    \label{EnthalpyMixCuNiSolid}
				\end{figure}

\begin{figure}[h]
        
                \centering
				\includegraphics[width=.81\linewidth]{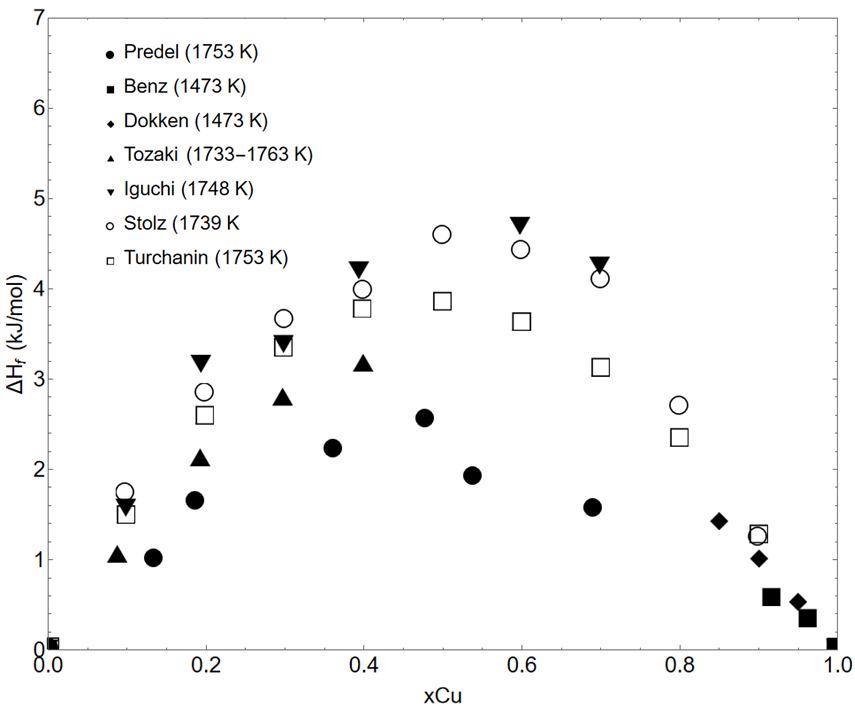}
				\caption{Enthalpy of mixing measured through calorimetric and EMF methods for liquid Cu-Ni alloys\cite{predel1971thermodynamic,benz1963high,dokken1965calorimetry,tozaki1971ban,iguchi1977calorimetric,stolz1993determination,turchanin1996heat}.}
    \label{EnthalpyMixCuNiLiquid}
				\end{figure}

Prigogine and Defay define an ideal associated solution as those for which the heat of mixing is either negligible (athermal) or of the same magnitude as the thermal energy (Regular solutions).  Here we will discuss solutions where the heat of mixing attains much larger values.  Figure (\ref{EnthalpyMixCuNiSolid}) plots the enthalpy of mixing for solid Cu-Ni alloys.  Despite some variation in the experimental measurements,  the broad trends are that $\Delta$H$^{\text{mix}}$ is large,  positive,  and askew.  With such large enthalpies of mixing, it is not likely that configurational entropy is at all similar to Equation (\ref{configurationalentropy}).  Similar trends are reported for experimental data  in the liquid state,  as shown in Figure (\ref{EnthalpyMixCuNiLiquid}).
Positive enthalpies of mixing encourage unmixing of the alloy system and suggest a preference of atoms for the same-type within their local atomic environments.  One can then begin to consider how allowing for the self-association of atoms into clusters (also called associates or complexes) alters the thermodynamics of mixing. 

In the case where only $n_\text{A}$ atoms of A are associating into clusters made of $i$ atoms;
\begin{equation}
n_{\text{A}} = \sum_{i} i n_{A_i}
\end{equation}
where $n_{A_i}$ is the number of such clusters,  and one possible cluster size among all clusters may be represented by $A_i$.  The clusters are in equilibrium with themselves and monoatomic $A_1$ and $B_1$. Neglecting for the moment enthalpic interactions among the clusters themselves,  which would otherwise result in simply a larger cluster,  one obtains an ideal associated solution (IAS).  The chemical potentials and concentrations for $A_1$ and $B_1$ in this model are then given by 

\begin{equation}
\mu_{A_1} = \mu_{A_1}^{\circ}(T,P) + RT\ln x_{A_1}
\end{equation}

\begin{equation}
\mu_{B_1} = \mu_{B_1}^{\circ}(T,P) + RT\ln x_{B_1}
\end{equation}

Herein,  in addition to the self association of the end members,  we include the existence of a mixed A-B clusters. To evaluate the thermodynamic potentials and quantities for such cluster model,   three aspects of the clusters are needed;  How many atoms are in a given  cluster,  how many bonds are within the cluster,  and what fraction of those clusters exist. 

\begin{figure}[h]
        
                \centering
				\includegraphics[width=.81\linewidth]{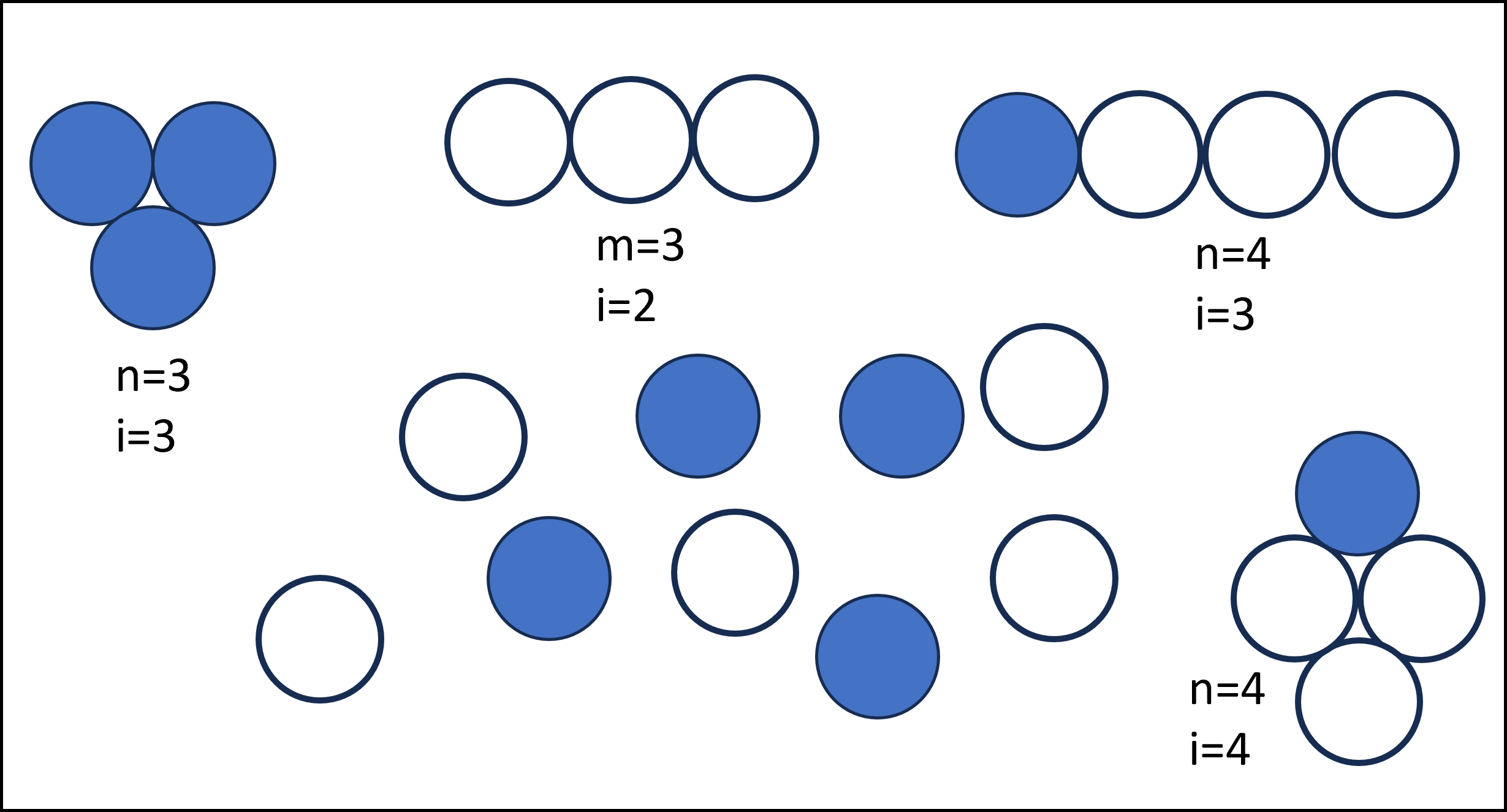}
				\caption{Illustration of some fo the relationship between the number of atoms in a cluster, $n$, and the number of bonds $i$. This Figure also illustrates how the number of bond types in the case of $A_{3}B$ clusters can be different based on topological changes.}
    \label{ClusterFigure}
				\end{figure}

More recent work by Shunyaev et al,  in the same vain as Prigogine,  treat the equilibrium between monatomic A and B and their clusters as chemical reactions \cite{shunyaev2002arbitrary}.  The fraction of a given cluster is $x_{A_{n,i}}$),  where $A$ refers to the atom type,  $n$,  refers to the number of atoms in the cluster, and $i$ the number of bonds.  Figure \ref{ClusterFigure} illustrates several of such clusters.  The law of mass action constrains the association of clusters in the melt (and in the solid, for justification, see \cite{kikuchi1977vi}) as a function of the fraction of monatomic atoms in solution ($x_{A_1}$:

\begin{equation}
x_{A_{n,i}} = K_{A_{n,i}} x_{A_1}^n = \exp \left(\frac{-\alpha_{A}i}{kT}\right) x_{A_1}^n
\end{equation}

\begin{equation}
x_{B_{n,i}} = K_{B_{n,i}} x_{B_1}^n = \exp \left(\frac{-\alpha_{B}i}{kT}\right) x_{B_1}^n
\end{equation}

\begin{align}
x_{A_n B_m} &= K_{A_{n,i}B_{m,j}} x_{A_1}^n x_{B_1}^m \nonumber \\
&= \exp \left(\frac{\alpha_{A}i+\alpha_{B}j-\alpha_{A_{n}B_{m}}q}{kT}\right) x_{A_1}^n x_{B_1}^m
\end{align}

Where $n$ or $m$ refers to the number of atoms in a particular cluster and $i$, $j$, and $q$ are the number of a particular type in a cluster. In the case of our model, we have used a cluster of stoichiometry $A_{3}B$. We make two critical assumptions that must be pointed out. We have not specified the relationship between $n$, the number of atoms in a cluster, and $i$, the number of bonds. One can imagine,  as illustrated in Figure\ref{ClusterFigure} a cluster of 3 atoms could be arranged in a triangle,  with 3 bonds,  or in a linear chain, with 2 bonds \cite{tkachev1989configurational}.  Work by Tkachev and others have explored how including topology alters the results but for the purposes of this article, we assume that the self-associating clusters of like-type atoms form linear chains. This implies that $i=n-1$. This makes the sum of the various associate compositions mathematically tractable and sufficient for qualitative explanation of the configurational entropy curve with mixed and self-associates \cite{shunyaev2002arbitrary}. In the case of the mixed associate $A_{3}B$ we include a single 4 atom cluster without a sum over the number of atoms in the equivalent stoichiometry (i.e.  8, 12, 16, etc...). The bonding parameter we used herein ($\alpha_{Cu_{3}Ni}q = -45$ kJ/mol) subsumes the number of bonds within the cluster itself. 

One then solves these equations for $x_{A_1}$ and $x_{B_1}$ to evaluate the chemical potential of A and B in  the solution.  We apply a normalization constraint for the mole fraction of clusters and atoms as:

\begin{equation}
\sum_{n, i} (x_{A_{n,i}} + x_{B_{n,i}}+x_{A_{n}B_{m}}) = 1 
\end{equation}

and the law of mass action:

\begin{equation}
\frac{\sum_{n, i} (nx_{A_{n,i}}+nx_{A_{n}B_{m}})}{\sum_{n, i} (nx_{A_{n,i}} + nx_{B_{n,i}}+(n+m)x_{A_{n}B_{m}})} = c_{A}. 
\end{equation}

Where $c_{A}$ in this instance is related to the oveall composition fraction of A ($X_{\text{A}}$) not to be confused with the fraction monatomic A atoms in solution.  Applying the assumption $i = n-1$ complete the equations and boundary conditions for the model.  We shall now examine its consequences.

The chemical potential of A in this solution model is then given by 

\begin{equation}
\mu_{\text{A}}^{\text{M}} = RT\ln \left(\frac{x_{A_{1}}}{x_{A_{1}}^{\circ}}\right),
\end{equation}

and by symmetry for B

\begin{equation}
\mu_{ \text{B} }^{ \text{M} } = RT\ln \left(\frac { x_{B_{1} } } { x_{B_{1}}^{\circ}}\right),
\end{equation}

where $x_{A_{1}}^\circ$ is the fraction of isolated atoms in the pure liquids of A (resp B) with the subscript indicating n=1 atom per cluster with i = 0 bonds and the superscript denoting self-association in the pure state rather than in the presence of other alloying elements.  This is needed to account for some amount of self-association in any pure system where  these clusters may exist.  It implies to solve for this fraction of isolated atoms, and changing our reference state from the chemical potential of isolated pure atoms to the chemical potential of the self-associated A. This manoeuvre is necessary because the reference state of the ideal associated system is for monatomic species; this change of reference state allows obtain thermodynamic partial and integral quantities that are referenced to the Raoultian state \cite{prigogine1958chemical}, a common practice for binary solutions A-B across the entire composition range.

The binding energy must now be proscribed for self association. Many empirical relationships have been provided to for the self-associate binding energy,  which can be either be endo or exothermic \cite{tkachev1989configurational}.  We chose a relationship from the literature for FCC metals given by the Equation:

\begin{equation}
\alpha_{(\text{A or B})} = 0.177RT_{m}^{(\text{A or B})}
\end{equation}

where R is the ideal gas constant and $T_{m}^{(\text{A or B})}$ is the melting point of A or B.  The skewmorphic entropy of mixing shown in Figure (\ref{CuNiEEntroypMixing}) from the electronic component alone would not be currently rationalizable with the conventional thermodynamic measurement techniques and data for Cu-Ni.  We therefore appeal to a cluster model here to alter the configurational entropy given the context of the measured electronic entropy produced in this report. Much of the expected behavior, such as a minimum in the configurational entropy of mixing of a mixed associate Cu$_{3}$N can be shown to emerge from such simple set of assumptions. 

\section{X-Ray Diffraction}
The samples were established to be single-phase after annealing and processing by XRD of the bulk bars in a Bragg-Bratano geometry with Cu-K$\alpha$ radiation. 

\begin{figure*}[h]
        
                \centering
				\includegraphics[width=.8\linewidth]{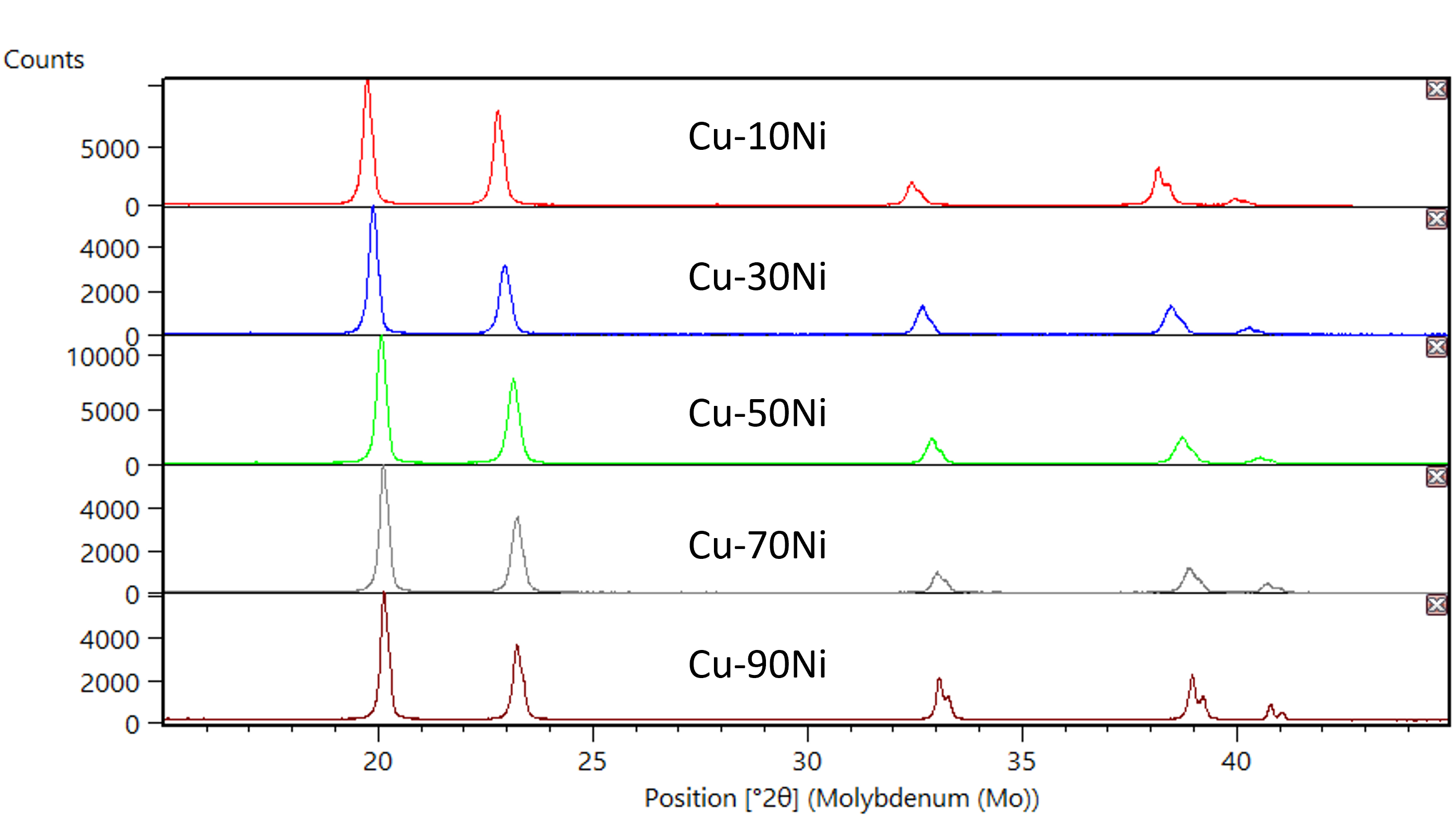}
				\caption{XRD Measurements of odd compositions of Cu-Ni alloys}
    \label{CuNiXRD}
				\end{figure*}
    
\begin{figure*}[h]
                \centering
				\includegraphics[width=.8\linewidth]{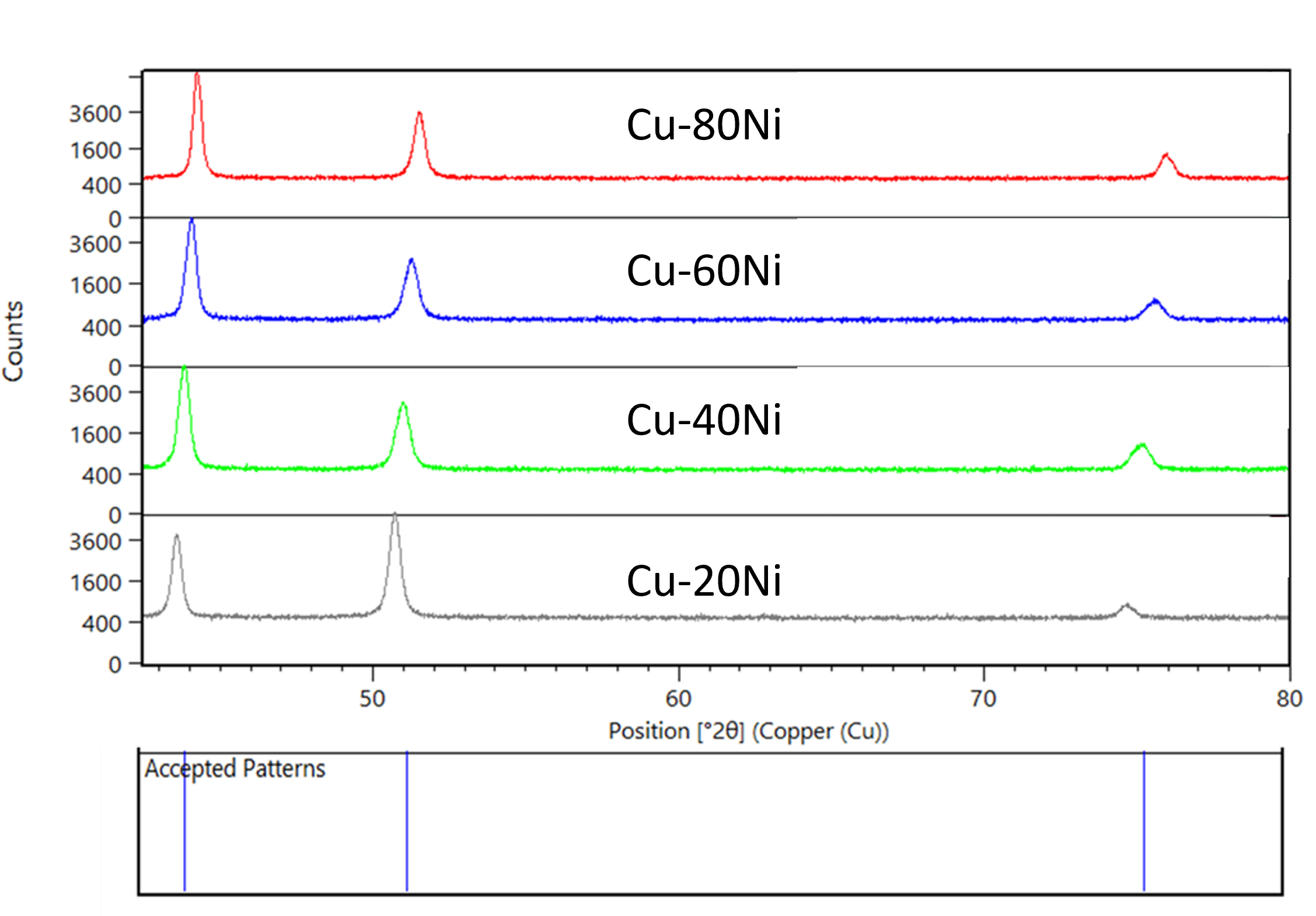}
				\caption{XRD Measurements of even compositions Cu-Ni alloys}
    \label{CuNiXRDEven}
				\end{figure*}
    
\section{Previous Thermodynamic Modeling Efforts}

The cluster model we will apply is similar to the one in Appendix B without a mixed associate to illustrate how such models may seem to correctly model the activity coefficient, with incorrect enthalpy and entropies of mixing. 
				\begin{figure*}[h]
                
                \centering
				\includegraphics[width=1.\linewidth]{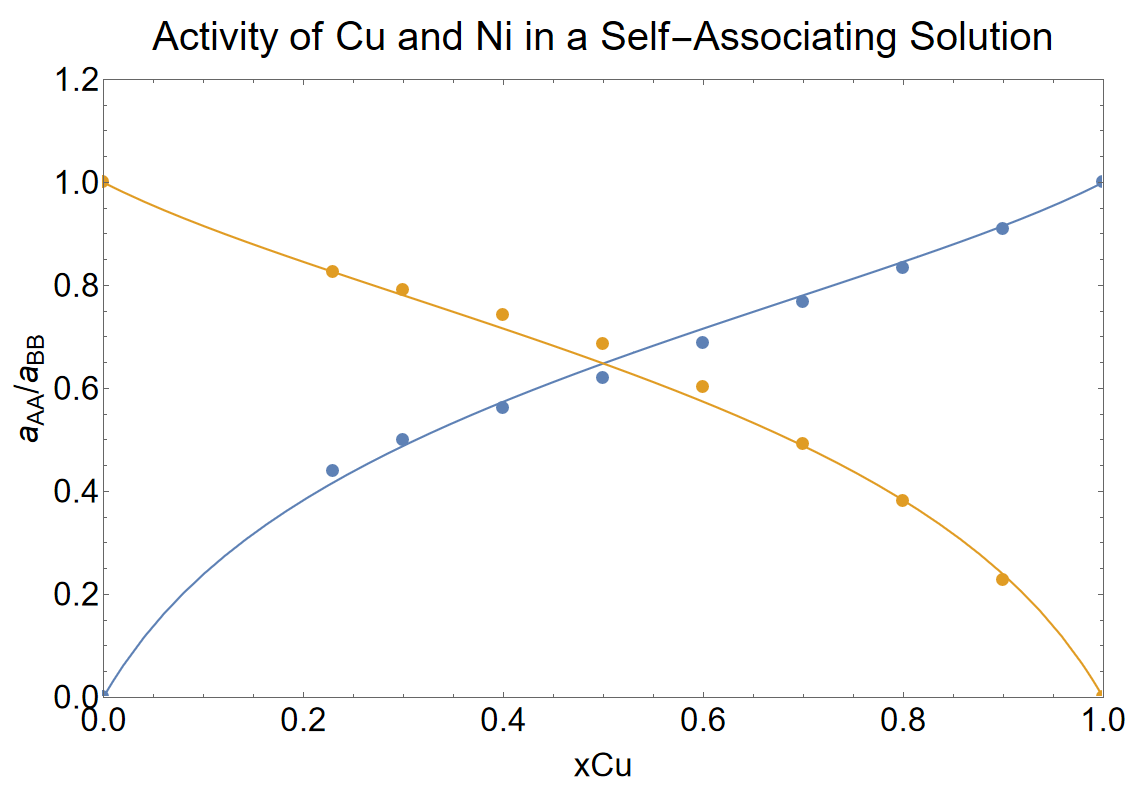}
				\caption{Implementation of an ideal associated solution model to modeling (lines) the activity coefficient of Cu-Ni in the liquid state in comparison to the data of Kulkarni (dots)\cite{Kulkarni1973}.}
    \label{ActivityCoefficient}
				\end{figure*}

While it may seem that the activity is well-modeled by such a system, we shall now present figures of the entropy and enthalpy of mixing, which we will see result in a reduced entropy (which is reassuring) but the wrong sign and magnitude of the enthalpy of mixing. 

\begin{figure*}[h]
                
                \centering
				\includegraphics[width=1.\linewidth]{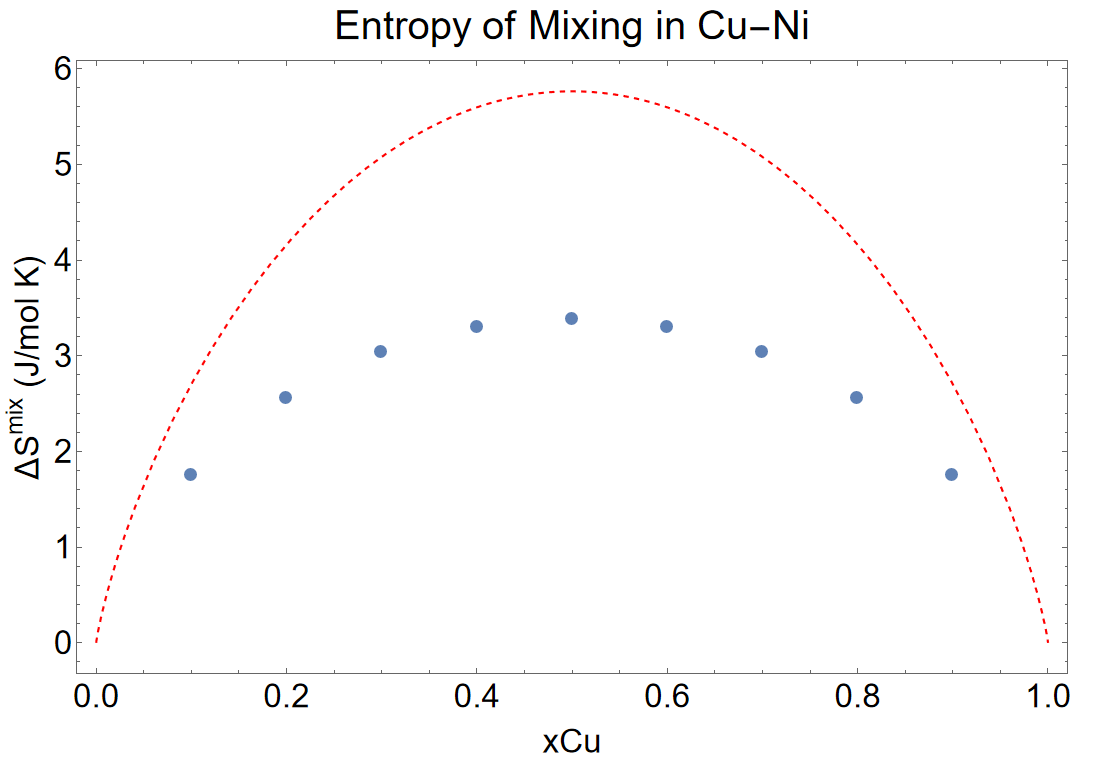}
				\caption{Implementation of an ideal associated solution model to the entropy of mixing (blue dots) with the uncorrelated configurational entropy of mixing plotted in red.}
    \label{EntropyOfMixingIdealSolutionModel}
				\end{figure*}

\begin{figure*}[h]
                
                \centering
				\includegraphics[width=1.\linewidth]{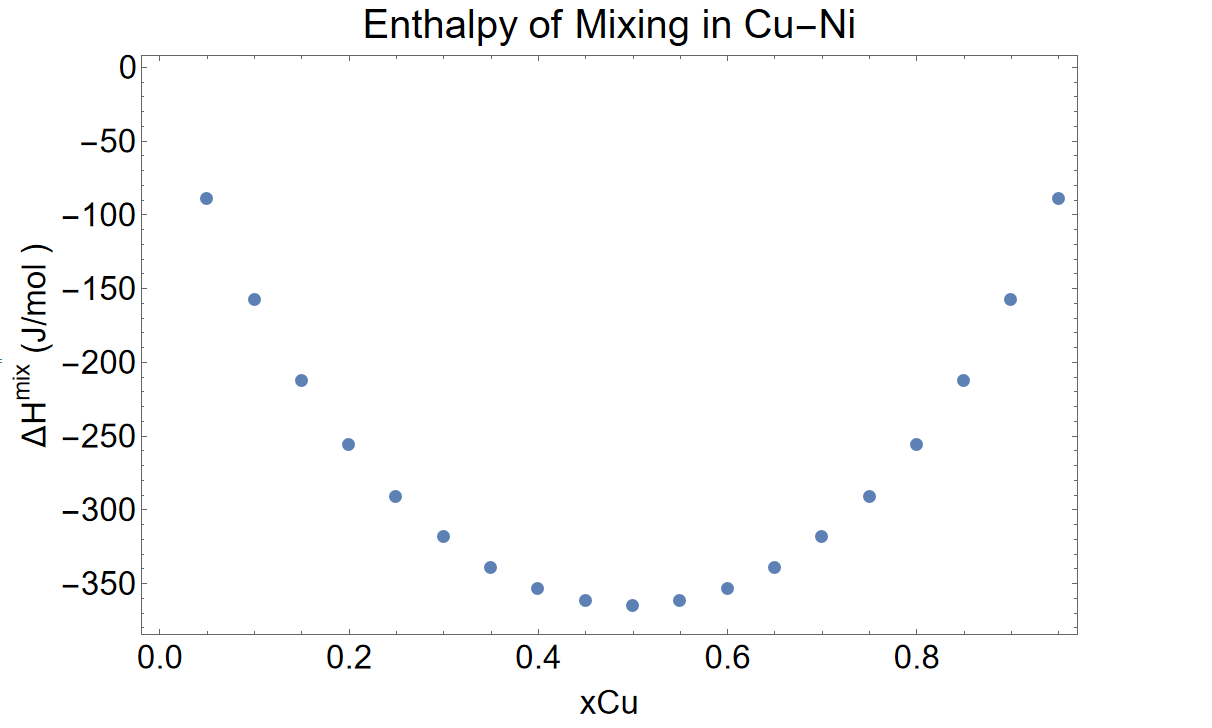}
				\caption{Implementation of an ideal associated solution model to the enthalpy of mixing (blue dots)}
    \label{EnthalpyOfmixingIAS}
				\end{figure*}

Figure (\ref{ActivityCoefficient}) demonstrates that we can reproduce the measured activity coefficient quite well using such a model for the system. However, the entropy and enthalpy of mixing (Figures (\ref{EnthalpyOfmixingIAS}) and (\ref{EntropyOfMixingIdealSolutionModel})) in this case, are both wildly different than their experimentally measured values. This is because the IAS does not have an enthalpy of mixing term other than the cluster formation energies. Such corrections can be applied ad-hoc to repair the enthalpy of mixing, but then revision of the cluster energies and the existence of certain clusters becomes necessary because the Gibbs energy and entropy are now wrong at the integral level.


\end{document}